\documentclass[superscriptaddress,aps,prx,onecolumn,amsmath,amssymb,11pt]{revtex4-1}
\usepackage{epsfig}
\usepackage{graphicx}
\usepackage{dcolumn}
\usepackage{subfigure}
\usepackage{hyperref}
\usepackage{ulem}
\usepackage{bm}
\usepackage{color}
\usepackage[mathscr]{euscript}
\usepackage{latexsym}
\hypersetup{bookmarks,
            colorlinks,
           citecolor=red,
            filecolor=blue,
           urlcolor=blue,
           pdfpagemode=None}
\usepackage{latexsym}
\def \k {{\bf k}}

\def \e {\varepsilon}
\def \r {{\bf r}}
\def \q {{\bf q}}
\def \Q{{\bf Q}}
\def \K{{\bf K}}

\def \D{\Delta}

\def \v{{\bf v}}
\def \ol{\overline}

\def \beq {\begin{eqnarray}}
\def \eeq {\end{eqnarray}}
\def \tn {\textnormal}
\begin{document}
\title{Feedback of superconducting fluctuations\\ on charge order in the underdoped cuprates}
\author{Debanjan Chowdhury}
\affiliation{Department of Physics, Harvard University, Cambridge Massachusetts 02138, U.S.A.}
\author{Subir Sachdev}
\affiliation{Department of Physics, Harvard University, Cambridge Massachusetts 02138, U.S.A.}
\affiliation{Perimeter Institute for Theoretical Physics, Waterloo, Ontario N2L 2Y5, Canada}
\begin{abstract}
Metals interacting via short-range antiferromagnetic fluctuations are unstable to sign-changing superconductivity at low temperatures. For the cuprates, this leading instability leads to the well-known $d-$wave superconducting state. However, there is also a secondary 
instability to an incommensurate charge density wave, with a predominantly $d$-wave form factor, arising from the same antiferromagnetic fluctuations. Recent experiments in the pseudogap regime of the hole-doped cuprates have found strong
evidence for such a charge density wave order and, in particular, the predicted $d$-wave form factor. 
However, the observed wavevector of the charge order differs from the leading instability in Hartree-Fock theory, and is that of a subleading instability. In this paper, we 
examine the feedback of superconducting fluctuations on these different charge-density wave states, and find that over at least a small temperature window, they prefer the experimentally observed wavevector. 
\end{abstract}
\maketitle
\section{Introduction} 
The pseudogap (PG) regime of the hole-doped cuprates is possibly one of the most enigmatic phases of matter. It has often been described as one of the central problems in the physics of high-temperature superconductivity (SC). It is identified by the onset of a large gap ($\sim 50$ meV) below a temperature, $T^*$, as observed in a number of different probes. The gap persists down to the superconducting transition temperature, $T_c$, below which of course the system develops the usual superconducting gap. The nature of the fermionic excitations in the PG phase is particularly interesting---the aforementioned gap is present in the anti-nodal regions of the Brillouin zone close to momenta $(\pi,0)$ and $(0,\pi)$; there exists, however, a region close to the nodes that remains gapless as has been detected via ARPES experiments \cite{ZXS03, ARCS}. These regions are commonly referred to as the fermi-``arcs". 

A topic of intense debate has been whether in addition to pairing fluctuations, any other possibly short-ranged and fluctuating phase with broken symmetry is also present above $T>T_c$, and below the onset of the pseudogap. Investigating the precise nature of this short-range order will likely shed some light on the ``normal" state out of which it emerges.
The different perspectives on this short-range order in the PG phase can be divided into two classes ({\it i\/}) the order
is `quantum disordered' and there are fractionalized excitations and associated topological excitations \cite{YRZ,Wen12,PS12,DCSS14}, or ({\it ii\/})
the order is ``classically disordered'' primarily due to thermal fluctuations. Here, we shall explore consequences arising from the second point of view. Implications of these thermal fluctuations on the phenomenology of the pseudogap phase, with comparison to various experiments, have also been explored recently \cite{LHSS14, AADCSS14}.

A huge leap in our understanding came about with the discovery of quantum oscillations caused by a small electron-like pocket at very large magnetic fields \cite{LT07}. This was the first clear evidence of the ``normal" state in the   pseudogap phase under large magnetic fields having some resemblance to a metallic state. More recently, in a series of remarkable experiments, it has become clear that the reconstructed electron-like pocket is caused by an incommensurate charge-density wave (CDW), competing with superconductivity \cite{Ghi12,DGH12,SH12,comin13, DGH13,MHJ11,MHJ13,CP13, JH02,AY04,JSD1011,comin2,SSJSD14,sonier}. The wavevector of this charge-density wave, which is of the type $(\pm Q_0,0),~(0,\pm Q_0)$, appears to be linked to the Fermi surface in the anti-nodal regions. Furthermore, diamagnetism measurements in YBCO show significant fluctuation diamagnetism over approximately the same range of temperatures where X-ray experiments measure charge order fluctuations, indicating that there are significant superconducting fluctuations in this phase \cite{Ong10, Cooper13}. A natural question that needs to be addressed is if both SC and CDW arise out of the same physics.

In the underdoped cuprates, due to the proximity to an antiferromagnet (AF) close to half-filling, the susceptibility, $\chi(\q)$, is peaked around $\K=\pm(\pi,\pi(1-\delta)),~\pm(\pi(1-\delta),\pi)$, ($\delta\neq0$ when the fluctuations are peaked around an incommensurate wavevector) \cite{Keimer10}. It is now well-understood that a metal interacting via such antiferromagnetic exchange interactions is unstable to $d-$wave superconductivity at low temperatures \cite{SCAL86, CMV86, SCAL10}. However, an interesting consequence of the AF exchange interactions is that they also give rise to a secondary instability to a charge-density wave with a predominantly $d$-wave
form factor \cite{MMSS10,SSRP13}. Significant recent developments have been the evidence for the predicted $d$-wave form factor
in X-ray observations \cite{comin2}, and a direct phase-sensitive measurement of the $d$-wave form factor in scanning tunneling microscopy (STM) experiments \cite{SSJSD14}. In the present paper we will turn our attention to the wavevector of this CDW, and 
in particular, its connection to the SC fluctuations.

\begin{figure}
\begin{center}
\includegraphics[width=0.4\columnwidth]{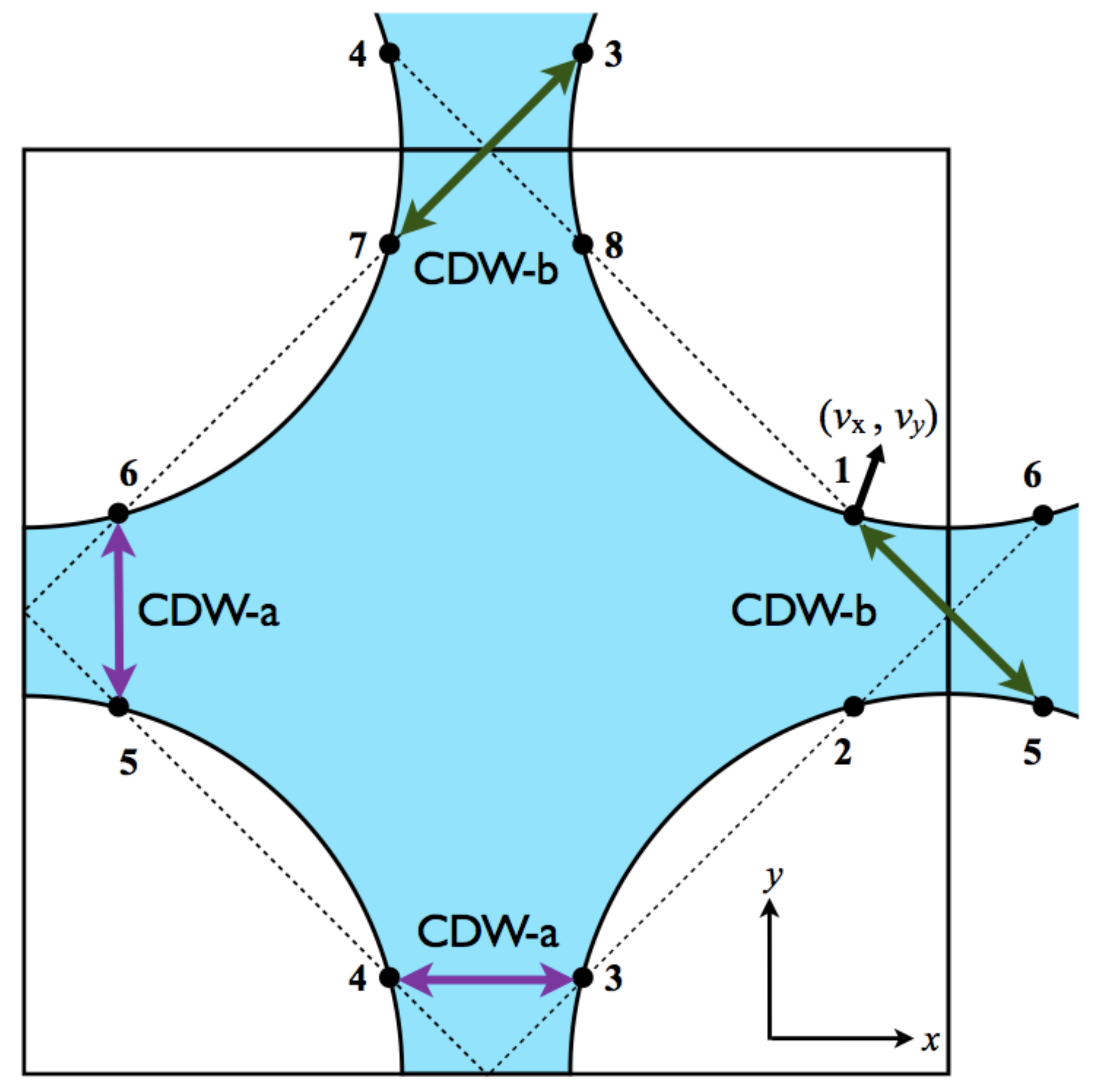}
\end{center}
\caption{Generic Fermi-surface for hole-doped cuprates, with the filled states shaded in blue. Solid circles, numbered $1,..,8$, represent hot-spots where the Fermi-surface intersects the magnetic Brillouin zone for a $(\pi,\pi)$ SDW. The purple arrows represent the CDW-a wavevectors, which closely resemble the experimentally observed wavevector, while the green arrows represent the CDW-b wavevectors, which arise as the leading instability in a HF calculation . The fermi-velocity, $(v_x,v_y)$, is shown at the hot-spot $1$. The wavevectors of CDW-a are $(\pm Q_0,0)$ and $(0,\pm Q_0)$, while those of CDW-b are $(Q_0, \pm Q_0)$. The computation in the present paper applies for a range of values of $Q_0$, and not just when it is equal to the separation between hot-spots as shown above; for different $Q_0$
we have to consider a corresponding set of 8 points around the Fermi surface, and our computations proceed unchanged. }
\label{bz}
\end{figure}

A number of recent works have addressed such issues \cite{MMSS10, SSRP13, SSJS14,DHL13,HMKE,HMKE13,HF14,AASS14,AASS14b,YWAC14,norman14,ATAC14,AKB14}.
In Refs.~\onlinecite{MMSS10,SSRP13}, the wavevector of the leading CDW instability has been found to be of the form $\pm(Q_0,Q_0),~\pm(Q_0,- Q_0)$, while the experimentally observed wavevector was a subleading instability. 
At this point it is useful to introduce some new notation. From now on, we shall often refer to the experimentally observed CDW with wavevectors $(Q_0,0)$ and $(0,Q_0)$ as ``CDW-a" and the one with wavevectors $(Q_0,\pm Q_0)$ as ``CDW-b" (see fig.\ref{bz}). 
More recently, it has been pointed out that in the presence of strong correlations arising from on-site and nearest-neighbor Coulomb repulsion, it is possible to obtain a stable CDW-a phase, with the wavevectors seen in experiments, in a certain window of parameter space  \cite{AASS14}. Wang and Chubukov \cite{YWAC14} examined retardation effects linked to the 
damping of spin fluctuations with a long correlation length, and argued for an extension of CDW-a with time-reversal symmetry breaking.
 In this paper, we will show that upon accounting for the interplay between superconductivity and charge-order in the underdoped cuprates, the CDW state with the correct wavevector ({\it i.e.\/} CDW-a) appears to be preferred over CDW-b over at least a small temperature window, as long as the antinodal regions of the fermi-surface are reasonably nested with a small curvature (a criterion to be made more precise later) \cite{commenttemp}. In fig.\ref{bz}, $Q_0$ represents the separation between two neighboring ``hot-spots". However, our computations in this paper are applicable for a range of different values of $Q_0$ that connect points in the antinodal regions and can be different from the hot-spot wavevector. Meier {\it et al.\/} \cite{HMKE13} have also recently looked at the effect of superconducting fluctuations on charge-order in a different setup.

We also note our recent work \cite{DCSS14}, which employs a “quantum disordered” model of the pseudogap
as a topological metal, and proposes an alternative mechanism for charge ordering at the 
($Q_0,0$), ($0,Q_0$) wavevectors.

Our paper is organized as follows: In section \ref{hf}, we introduce the theory of a metal interacting via antiferromagnetic exchange interactions (henceforth referred to as  $t-J$ model, but without any on-site Coulomb repulsion) and review a Hartree-Fock analysis for various charge-ordering and pair-density wave instabilities. Based on the leading instabilities that occur in a metal, we construct the minimal model in section \ref{mod} for a metal with pairing and charge-order fluctuations. In section \ref{let}, we take the low-energy limit of this theory and present the effective theory in the vicinity of special points---the ``hot-spots"---where the magnetic Brillouin-zone corresponding to $\K$ intersects the fermi-surface. We present the results of our computation, describing the mutual feedback of superconductivity and charge-order on each other, in sections \ref{reshs} and \ref{rescu}. Finally in section \ref{dis}, we summarize the main results emerging from this analysis. Some of the technical details are summarized in appendices \ref{int} and \ref{intcurv}.

\section{Model} 
\subsection{Metal with antiferromagnetic exchange interaction}
\label{hf}
In this section, we briefly review some of the earlier results \cite{SSRP13} obtained by carrying out a Hartree-Fock (HF) analysis of the t-J model (without any Gutzwiller projection). Consider the following model for fermions,  $c_{i\alpha}$, interacting via short-range antiferromagnetic exchange interactions,
\beq
H_{tJ}=\sum_{i,j}\bigg[(-t_{ij}-\mu\delta_{ij})c_{i\alpha}^\dagger c_{j\alpha} + \frac{1}{2}J_{ij}\vec{S}_i\cdot\vec{S}_j\bigg],
\eeq
where $t_{ij}$ are the hopping amplitudes, $\mu$ denotes the chemical potential, $J_{ij}$ are the AF exchange couplings and $\vec{S}_i=c_{i\alpha}^\dagger\vec{\sigma}_{\alpha\beta} c_{i\beta}/2$.

We are interested in looking at the various instabilities that can arise in this model in the particle-particle as well as particle-hole channel. We shall restrict our attention to pair-density wave (PDW), where the SC condensate carries a finite momentum, and charge-density wave states. We ignore the possibility of having spin-order, partly motivated by most experiments on the non-La-based cuprates (e.g. YBCO and BSCCO), where the region of charge-order has hardly any overlap with that of spin-order \cite{MHJ11}. For the Hartree-Fock analysis, we need the best variational estimate for the following mean-field Hamiltonian,
\beq
H_{MF}=\sum_\k \bigg[\e(\k)c_{\k,\alpha}^\dagger c_{\k,\alpha} &+& \sum_\Q \D_Q(\k) \epsilon_{\alpha\beta}c_{-\k+\Q/2,\alpha}c_{\k+\Q/2,\beta} \nonumber\\
&+& \sum_\Q P_Q(\k)c_{\k+\Q/2,\alpha}^\dagger c_{\k-\Q/2,\alpha} + \tn{H.c.}\bigg].
\eeq
In the above, $\e(\k)$ represents the electronic dispersion. All the functions, $\D_Q(\k), P_Q(\k)$ are variational parameters which will be optimized by minimizing the free energy,  $F\leq F_{MF}+\langle H-H_{MF}\rangle_{MF}$. As mentioned earlier, we have allowed for a spin-singlet pair-density wave along with a charge density wave at a finite wavevector, $\Q$. The pair-density wave at $\Q\rightarrow0$ reduces to the standard BCS state where particles with opposite spins are paired at $\pm\k$. 

 Expanding the right-hand side of $F$ in powers of $\D_Q(\k)$ and $P_Q(\k)$, we get,
\beq
F&=&2\sum_{\k,\k',\Q}\D_Q^*(\k)\sqrt{\Pi_S(\k)} {\cal{M}}_S(\k,\k')\sqrt{\Pi_S(\k')}\D_Q(\k') \nonumber\\
&+& \sum_{\k,\k',\Q}{P}_Q^*(\k)\sqrt{\Pi_C(\k)} {\cal{M}}_C(\k,\k')\sqrt{\Pi_C(\k')}{P}_Q(\k'),
\label{FSC}
\eeq 
where the kernels ${\cal{M}}_{S,C}(\k,\k')$ are given by,
\beq
{\cal{M}}_{S,C}(\k,\k')=\delta_{\k,\k'}+\frac{3}{V}~\chi(\k-\k')\sqrt{\Pi_{S,C}(\k)\Pi_{S,C}(\k')},
\eeq
and the polarizabilities are,
\beq
\Pi_S(\k)&=& \frac{1-f(\e(\k+\Q/2))-f(\e(-\k+\Q/2))}{\e(\k+\Q/2)+\e(-\k+\Q/2)},\\
\Pi_C(\k)&=& \frac{f(\e(\k+\Q/2))-f(\e(\k-\Q/2))}{\e(\k-\Q/2)-\e(\k+\Q/2)},
\eeq
with $f(...)$ the Fermi-function and $\chi(\k)(=\sum_{i,j}J_{ij}e^{i\q.\cdot(\r_i-\r_j)}/4)$, the AF susceptibility. Note that for dispersions that satisfy $\e(\k+\Q)=-\e(\k)$, $\Pi_S(\k)=\Pi_C(\k)$. This holds in the vicinity of the hot-spots for certain values of $\Q$ and will have important consequences, which we shall revisit later. 

Pair-density wave and charge-ordering in the metal occurs via condensation in the eigenmodes of the operators ${\cal{M}}_{S,C}$ with the lowest eigenvalues. In order to find the leading instability in the pairing and charge-ordering channel, we need to solve the following eigenvalue problem,
\beq
\frac{3}{V}\sum_{\k'}\sqrt{\Pi_S(\k)}~\chi(\k-\k')\sqrt{\Pi_S(\k')}\phi_S(\k') &=& \lambda_S \phi_S(\k'),\\
\frac{3}{V}\sum_{\k'}\sqrt{\Pi_C(\k)}~\chi(\k-\k')\sqrt{\Pi_C(\k')}\phi_C(\k') &=& \lambda_C \phi_C(\k').
\label{eig}
\eeq

Instead of working with the form of $\chi(\k)$ introduced above, we assume that the AF susceptibility, $\chi(\q)$, has the form,
\beq
\chi(\q)=\sum_{\K}\frac{\chi_0}{4(\xi^{-2}+2(2-\cos(q_x-K_x)-\cos(q_y-K_y)))},
\eeq
which is peaked near the antiferromagnetic wavevector, $\K$ (as introduced earlier), with $\xi$ representing the AF correlation length and $\chi_0$ the overall strength of the spin-fluctuations. There is little difference between the results for $\delta=0$ and $\delta\neq0$.

\begin{figure}
\begin{center}
\includegraphics[width=0.4\columnwidth]{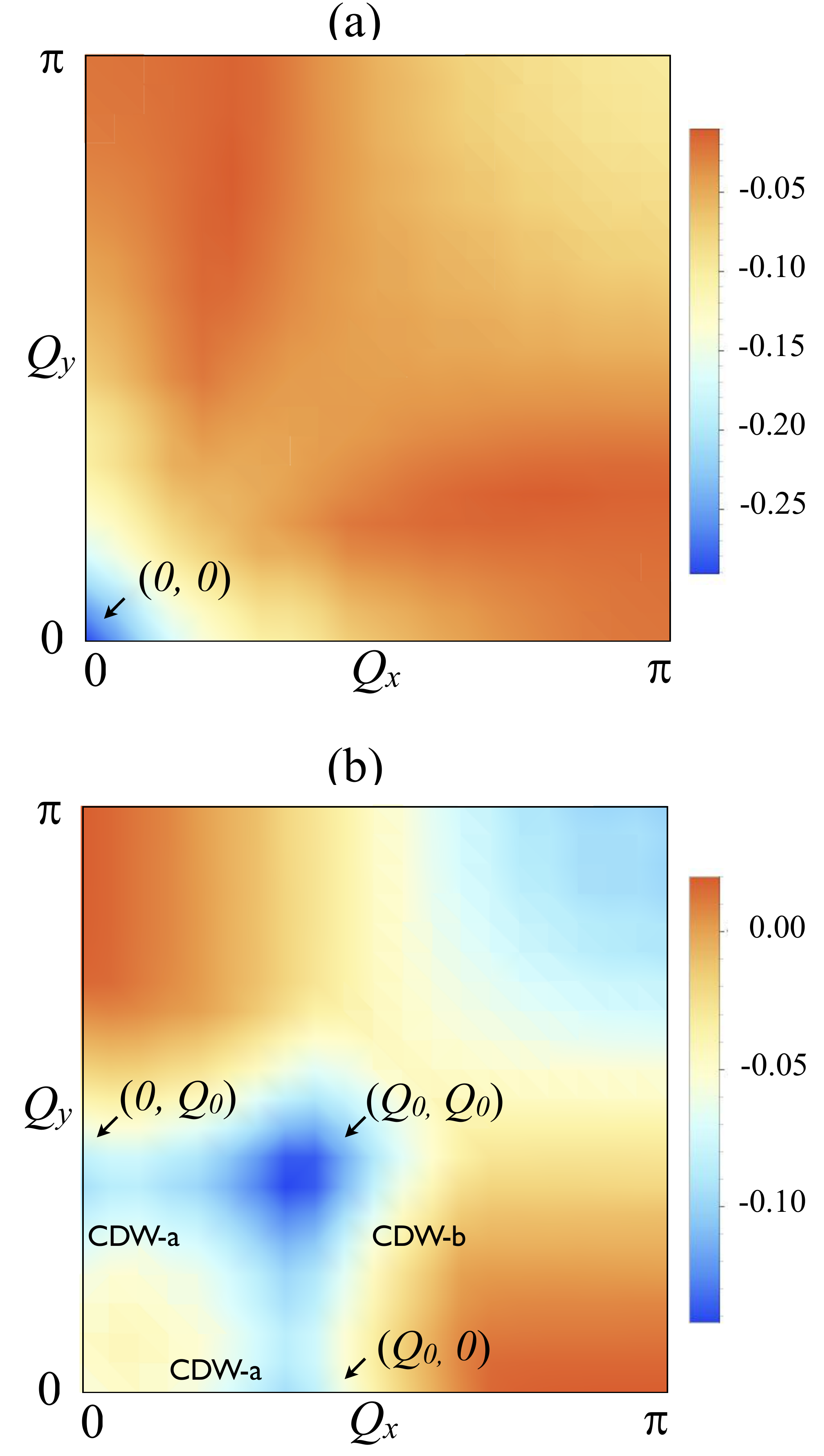}
\end{center}
\caption{Plot of the smallest eigenvalue (a) $\lambda_S/\chi_0$ and (b) $\lambda_C/\chi_0$ as a function of $Q_x$ and $Q_y$. We use a band-structure with $t_1=1.0,~t_2=-0.32,~t_3=0.128,~\mu=-1.11856$ (corresponding to the same fermi-surface used in Ref.\cite{SSRP13}). The other parameters are $\xi=2,~T=0.1,~\delta=1/4$ and $L=20$. For the pair-density wave, the global minimum is located at $\Q=0$ while for the charge-order, it is located at $\Q=(Q_0,Q_0)$ (CDW-b). In (b), note the valleys extending from $(Q_0,Q_0)$ to $(Q_0,0)$ and $(0,Q_0)$ (CDW-a)---the latter are local, but not global, minima.}
\label{eigfig}
\end{figure}

In figs. \ref{eigfig} (a) and (b) we plot the lowest eigenvalues $\lambda_{S,C}/\chi_0$, obtained after diagonalizing eqns.\ref{eig} on a discrete Brillouin zone with $L^2$ points. We use the following electronic dispersion: $\e(\k)=-2t_1(\cos(k_x)+\cos(k_y))-4t_2\cos(k_x)\cos(k_y)-2t_3(\cos(2k_x)+\cos(2k_y))-\mu$.

Let us start with a discussion of the pair-density wave state (fig.\ref{eigfig}a). Not surprisingly, the state with $\Q=0$ is the leading instability and in particular, $-\lambda_S$ for $\Q=0$ has a logarithmic divergence as $T\rightarrow0$. We do not find any other local-minima in phase-space; the BCS-log for $\Q=0$ simply overwhelms any other PDW state that might have otherwise arisen within this weak-coupling approach.  Comparing the numerical values of the eigenvalues $\lambda_S$ and $\lambda_C$ in figs.\ref{eigfig} (a) and (b), it is clear that $-\lambda_S$ (at $\Q=0$) is the smallest. Therefore, in the presence of short-range AF interactions, the leading weak-coupling quadratic instability indeed turns out to be to a SC (with $d-$wave symmetry; this information is contained in the structure of $\phi_S$). 

For the charge-ordering instability (fig.\ref{eigfig}b), the global minimum is located at the diagonal wavevector $(Q_0,Q_0)$. This corresponds to the CDW-b state introduced earlier. However notice the ``valleys" of local minima extending from this wavevector to $(Q_0,0)$ and $(0,Q_0)$, the CDW-a.  It is also worth pointing out that  the form factors associated with these CDWs, $\phi_C$, primarily have a $d-$wave component. In real space, this implies that the charge-density modulation is mostly on the Cu-Cu links (bonds) rather than on the sites. Hence they are also referred to as bond-ordered states. 

The reason why CDW-b turned out to be the leading CDW instability is actually related to two features in our problem above: {\it (i)} the absence of a gap in the spectrum at the antinodes, and, {\it (ii)} an {\it emergent} particle-hole symmetry associated with the exchange interactions, $J_{ij}\vec{S}_i\cdot\vec{S}_j$. This exchange interaction is invariant under independent SU(2) rotations on each lattice site that map particles to holes and vice-versa. These rotations therefore naturally map SC to CDW-b. In the absence of any fermi-surface curvature or other interaction terms that break this symmetry, both of these instabilites would be exactly degenerate. However, once it is broken, CDW-b becomes a sub-leading instability (SC is always guaranteed to remain the leading instability even in the presence of a curvature since the points $\pm\k$ are always ``nested". However, a very large nearest neighbor Coulomb interaction term, $Vn_in_j$, can suppress superconductivity, for instance). We also note in passing that CDW-a is mapped to the PDW \cite{PAL14} at the same wave-vector under these rotations; we do not consider this
PDW here because it does not appear as a preferred eigenvalue of $\lambda_S$ in fig.~\ref{eigfig}a.

The above analysis was carried out in the limit where the different orders were decoupled from each other, i.e. we did not look at the feedback of one over the other and analyzed the free-energy $F$ to only quadratic order. Let us therefore now construct a minimal model for SC and CDW fluctuations in a metal and analyze the effective theory beyond quadratic order in the next section.

\subsection{Metal with SC and CDW fluctuations}
\label{mod}
Consider now a model for the fermions with a generic fermi-surface as shown in fig.\ref{bz}. Based on our HF analysis in the previous section, we know the (sub-)leading instabilities were to a (i) $d-$wave superconductor, (ii) CDW-b with wavevectors $(Q_0,\pm Q_0)$, and, (iii) CDW-a with wavevectors $(Q_0,0)$ and $(0,Q_0)$. We can use this information to construct a model for a metal with strong pairing and CDW fluctuations. Therefore, we consider the theory for the fermions coupled to superconducting ($\Psi$) and CDW ($\Phi$) fluctuations (which are in their uncondensed phase, as is the case in the PG regime). The Hamiltonian is then given by,
\beq
H&=&H_0+H_S+H_B,~\tn{where}\\
H_0&=&\sum_{\k}(\e_\k-\mu)~ c_{\k,\alpha}^\dagger c_{\k,\alpha},\\
H_S&=&\sum_\k(\D_s(\k)~c^\dagger_{\k+\q/2,\uparrow}c^\dagger_{-\k+\q/2,\downarrow}\Psi_\q + \tn{H.c.}),\\
H_B&=&\sum_{\k,\q}\bigg[\bigg(\sum_\Q P_{Q}(\k)~ \Phi_{\q-\Q}\bigg) c_{\k+\q/2,\alpha}^\dagger c_{\k-\q/2,\alpha} + \tn{H.c.}\bigg].
\eeq
In the above, $\D_s(\k)$ is the usual form factor associated with $d-$wave superconductivity and $P_Q(\k)$ is the form-factor for CDW with wave-vector $\Q$. We would now like to obtain an effective action in terms of $\Psi$ and $\Phi$ (both CDW-a and b) after integrating out the fermions. The aim of this work is to study the effect of SC fluctuations on these two different CDW states and to analyze if the competition between the different order parameters can preferentially select a particular state. It will be particularly interesting if this state corresponds to the one that has been seen experimentally. 
 
\section{Low-energy theory} 
\label{let}
Instead of carrying out the above task with the full Fermi-surface, let us analyze the theory in the vicinity of the hot-spots labelled $j=1,..,8$ in fig.\ref{bz}. We expand the dispersion close to the hot-spots so that 
\beq
\e_{\k,j}=v_{F,\k_j}k_\perp + \kappa k_\Vert^2, 
\eeq
with $k_\perp (k_\Vert)$ being the momentum normal (parallel) to the Fermi surface. The Fermi-velocities are given by, $\v_{F,\k_1}=(v_x,v_y)$, $\v_{F,\k_2}=(v_x,-v_y)$, $\v_{F,\k_4}=(-v_y,-v_x)$ and $\v_{F,\k_7}=(-v_y,v_x)$. The other velocities can be obtained similarly by symmetry. The parameter $\kappa$ is related to the Fermi surface curvature. 

The bare Lagrangian for the fermions in the vicinity of the hot-spots, $\psi_j$ ($j=1,..,8$), is then given by,
\beq
{\cal{L}}_0&=&\sum_{j=1}^8\bigg[\psi_j^\dagger(i\omega-\e_{\k,j})\psi_j\bigg].
\label{L0}
\eeq
The same expansion can be carried out for a set of any eight points in the antinodal regions, that are not necessarily connected by the hot-spot wavevectors.

\subsection{Form factor of the CDW}
There are two fundamental properties associated with the CDW orders --- the wavevector and the structure of the form factor. We have explored the wavevectors that can arise in our HF computation in section \ref{hf}, while the form factors associated with the different CDW orders for the full underlying fermi-surface were already computed in Ref.\cite{SSRP13}. In this section, we shall revisit this issue within our low-energy formulation.

If we go back to eqn.\ref{FSC} and focus only on the terms involving charge-order, we obtain
\beq
F_C=\sum_\Q\bigg[\sum_{\k}|P_Q(\k)|^2 \Pi_C(\k) + \frac{3}{V}\sum_{\k,\k'} \chi(\k-\k')\Pi_C(\k)\Pi_C(\k') P_Q^*(\k) P_Q(\k')\bigg].
\eeq
We now assume that $\chi(\k-\k')$ is peaked at $\k-\k'=\K=(\pi,\pi)$ and perform the integrations over $\k$ and $\k'$ in patches in the neigborhoods of the hot-spots that satisfy the above constraint. Furthermore, we assume that $P_Q(\k)$ can be treated as piecewise constant in the patches and treat $\chi$ as static and non-critical, i.e. $\chi\approx\chi_0/\xi^{-2}$. For CDW-b, let us focus on the hot-spot pairs $\{2,6\}$ and $\{7,3\}$ where $P_Q(\k)$ takes the values $\Upsilon^b_1$ and $\Upsilon^b_2$. Similarly, for CDW-a, we choose to focus on the pairs $\{1,2\}$ and $\{4,7\}$ where $P_Q(\k)$ takes the values $\Upsilon^a_1$ and $\Upsilon^a_2$. 

It is straightforward to see that for CDW-b, $\Pi_C(\k)$ evaluated in the patches $\{2,6\}$ and $\{7,3\}$ are equal to each other due to purely geometric reasons, i.e. $\Pi^b_1=\Pi^b_2=\Pi^b$, so that,
\beq
F_C\bigg|_b=\Pi^b \bigg[|\Upsilon^b_1|^2+|\Upsilon^b_2|^2 \bigg] + \frac{3\chi_0}{\xi^{-2}} (\Pi^b)^2 \bigg[\Upsilon^{b*}_1\Upsilon^b_2 + \Upsilon^{b*}_2\Upsilon^b_1\bigg].
\eeq
It is simple to diagonalize the above quadratic form and obtain the optimum linear combination of $\Upsilon^b_1$ and $\Upsilon^b_2$. For CDW-b, the eigenvector corresponding to the lower eigenvalue has a purely $d-$wave form. 

We can now do the same computation for CDW-a and we immediately find that $\Pi^a_1\neq\Pi^a_2$ (once again, for purely geometric reasons), so that,
\beq
F_C\bigg|_a=\Pi^a_1|\Upsilon^a_1|^2 + \Pi^a_2|\Upsilon^a_2|^2 + \frac{3\chi_0}{\xi^{-2}} \Pi^a_1\Pi^a_2\bigg[\Upsilon^{a*}_1\Upsilon^a_2 + \Upsilon^{a*}_2\Upsilon^a_1 \bigg]. \label{FCUpsilon}
\eeq
We can diagonalize the above quadratic form and 
find that the eigenvector corresponding to the lower eigenvalue contains a mixture of $d-$ and 
$s-$wave forms. 
It is important to note that there is an ambiguity in choosing the eigenvector of the quadratic form \cite{SSRP13,AASS14b}: we can 
rescale $\Upsilon^a_1, \Upsilon^a_2$ by different factors ({\it i.e.\/} perform a similarity transform) 
before diagonalizing the quadratic form, and then undo the similarity transform after the diagonalization. This modifies the eigenvectors
except when the lowest eigenvalue is zero. It was argued \cite{SSRP13,AASS14b} that the appropriate similarity transform is determined 
by looking at the structure of the particle-hole T-matrix, which leads to the requirement that the diagonal terms in the quadratic form have equal values.
In this manner, we find that the eigenvector with lower eigenvalue has $(\Upsilon^a_1, \Upsilon^a_2) \propto (1/\sqrt{\Pi^a_1}, - 1/\sqrt{\Pi^a_2} )$.

In order to estimate the difference between $\Pi_1^a$ and $\Pi_2^a$, we can do an explicit computation at $T=0$ and in the absence of a fermi-surface curvature, so that 
\beq
\Pi_1^a=\frac{2}{4\pi^2v_xv_y}\int_{-\Lambda_x}^{\Lambda_x}dx~\int_{0}^{\Lambda_y}dy~\frac{\theta(y-x)-\theta(-y-x)}{2y},
\eeq
where $\Lambda_{x,y}=v_{x,y}\Lambda$, $\theta(...)$ represents the heaviside-step function and we are integrating near the hot-spots in a momentum window $|\k|<\Lambda$, with $\Lambda$ a UV regulator. We are interested in the limit $\Lambda_y > \Lambda_x$ (since $v_y>v_x$ in the antinodal regions) and obtain,
\beq
\Pi_1^a=\frac{\Lambda}{2\pi^2v_y}\bigg[1+\log\bigg(\frac{v_y}{v_x} \bigg) \bigg].
\eeq
On the other hand, 
\beq
\Pi_2^a&=&\frac{2}{4\pi^2v_xv_y}\int_{0}^{\Lambda_x}dx~\int_{-\Lambda_y}^{\Lambda_y}dy~\frac{\theta(y+x)-\theta(y-x)}{2x},\\
\Pi_2^a&=&\frac{\Lambda}{2\pi^2v_y}.
\eeq
Therefore, we see that,
\beq
\Pi^a_1&=&\eta\Pi^a_2>0,~\tn{where}\\
\eta&=&1+\log(v_y/v_x).
\eeq
We then conclude from the discussion below Eq.~(\ref{FCUpsilon}) that the ratio of the $s$ 
to the remaining bond components in the form factor of CDW-a is
\beq
 \left| \frac{\Upsilon^a_1 + \Upsilon^a_2}{\Upsilon^a_1 - \Upsilon^a_2}\right| = 
\frac{\sqrt{\eta} - 1}{\sqrt{\eta} + 1},
\eeq
which can be quite small. We expect the aforementioned remaining component of the CDW to be $d$. (Although
the present hot-spot computation does not, strictly speaking, distinguish between $s'$ and $d$, demanding smooth
variation of the form factor in the anti-nodal region strongly prefers $d$). 

Wang and Chubukov \cite{YWAC14} also analyzed the form-factor of CDW-a by looking at the set of coupled CDW vertices, retaining the Landau damping terms in the Bosonic propagator. (In particular our $\eta\rightarrow1$ limit corresponds to $\varphi\rightarrow\pi/4$ in their notation.)

\subsection{Interplay of charge-order and superconductivity}

In section \ref{hf}, we saw that at quadratic order one doesn't obtain the CDW with the experimentally measured wavevector. Therefore, it is necessary to go to quartic order; our real interest in this section is to compute these terms and, in particular, their temperature dependencies.

Let us now write the full low-energy theory in terms of the (a) fermions, $\psi_j$, (b) CDW -a ($\Phi^a_x$, $\Phi^a_y$) with wave-vectors $\Q^a_x=(Q_0,0)$ and $\Q^a_y=(0,Q_0)$, (c) CDW-b ($\Phi^b_x$, $\Phi^b_y$) with wave-vectors $\Q^b_x=(Q_0,Q_0)$ and $\Q^b_y=(Q_0,-Q_0)$, and, (d) SC ($\Psi$):
\beq
{\cal{L}}&=&{\cal{L}}_0+{\cal{L}}_S+{\cal{L}}_B,\\
{\cal{L}}_S&=&\Psi(\psi^\dagger_1 \psi^\dagger_5 + \psi^\dagger_2 \psi^\dagger_6) -  \Psi(\psi^\dagger_7 \psi^\dagger_3 + \psi^\dagger_4 \psi^\dagger_8) + \tn{H.c.},\\
{\cal{L}}_B&=& \Phi_x^a(\psi^\dagger_6 \psi_1 + \psi^\dagger_5 \psi_2 - \psi^\dagger_3 \psi_4 - \psi^\dagger_8 \psi_7) \nonumber\\
&-&  \Phi_y^a(\psi^\dagger_1 \psi_2 + \psi^\dagger_6 \psi_5 - \psi^\dagger_3 \psi_8 - \psi^\dagger_4 \psi_7) \nonumber\\
&+& \Phi_x^b(\psi^\dagger_6 \psi_2 - \psi^\dagger_3 \psi_7) -  \Phi_y^b(\psi^\dagger_5 \psi_1 - \psi^\dagger_8 \psi_4) + \tn{H.c.},
\eeq
where we have suppressed the momentum and spin-index structure above and ${\cal{L}}_0$ was already expressed in eqn.\ref{L0}. While writing ${\cal{L}}_B$, we have ignored the possibility of having a small $s-$wave component in the form factors of $\Phi_{x,y}^a$
{\it i.e.\/} we have assumed $P_Q (\k) = \cos k_x - \cos k_y$. 
We also choose not to write any explicit coupling constants as they can be absorbed into the fields by a redefinition. In the low-energy limit, the patches 1-2-5-6 and 3-4-7-8 are decoupled from each other. 

\begin{figure*}
\begin{center}
\includegraphics[width=1.0\columnwidth]{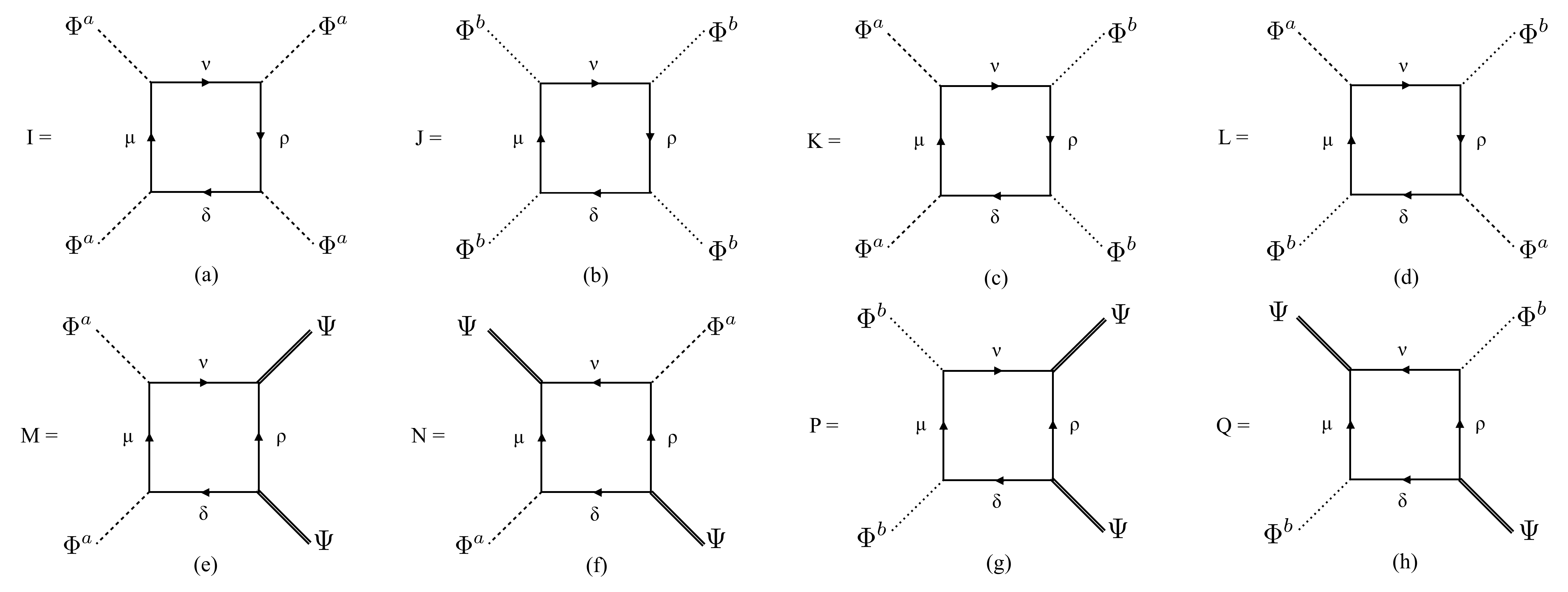}
\end{center}
\caption{Feynman diagrams representing the various 4-point functions that contribute to different terms in $S_{\tn{eff}}$. The solid internal lines carry different hot-spot indices ($\mu,\nu,\rho,\delta$) depending upon the term being evaluated. The dashed, dotted and double lines represent $\Phi^a$, $\Phi^b$ and $\Psi$ respectively. The individual diagrams are labelled (a) $I_{\mu\nu\rho\delta}$, (b) $J_{\mu\nu\rho\delta}$, (c) $K_{\mu\nu\rho\delta}$, (d) $L_{\mu\nu\rho\delta}$, (e) $M_{\mu\nu\rho\delta}$, (f) $N_{\mu\nu\rho\delta}$, (g) $P_{\mu\nu\rho\delta}$, (h) $Q_{\mu\nu\rho\delta}$. }
\label{fd}
\end{figure*}

Once again, we integrate out the fermions in the vicinity of the hot-spots (in a momentum window $|\k|<\Lambda$) and compute the action upto fourth order in $\Phi^a$, $\Phi^b$ and $\Psi$. All the four-point diagrams contributing to these terms are shown in fig. \ref{fd}. There is also a three-point diagram, as shown in fig.\ref{fdt}, contributing to the effective action, which takes the form,
\begin{widetext}
\beq
S_{\tn{eff}}[\Phi^a,\Phi^b,\Psi]=\int d^2\r ~d\tau &\bigg[&r_a\bigg(|\Phi_x^a|^2+|\Phi_y^a|^2\bigg) +  r_b\bigg(|\Phi_x^b|^2+|\Phi_y^b|^2\bigg) + r_b |\Psi|^2 \nonumber\\
&+&u_{a} \bigg(|\Phi_x^a|^4 + |\Phi_y^a|^4\bigg) +  u_{b} \bigg(|\Phi_x^b|^4 + |\Phi_y^b|^4\bigg) +  u_b|\Psi|^4\nonumber\\
&+& u_{ab} \bigg(|\Phi_x^a|^2 + |\Phi_y^a|^2\bigg) \bigg(|\Phi_x^b|^2 + |\Phi_y^b|^2\bigg) + w_a |\Phi_x^a|^2 |\Phi_y^a|^2 \nonumber\\
&+& \ol{u}_{ab}\bigg((\Phi_x^a)^2\Phi_x^{b*}\Phi_y^{b*} + (\Phi_y^a)^2\Phi_x^{b*}\Phi_y^b + \tn{H.c.}\bigg) \nonumber\\
&+& t_{ab}\bigg(\Phi_x^a\Phi_y^{a}\Phi_x^{b*} + \Phi_x^a\Phi_y^{a*}\Phi_y^{b*} + \tn{H.c.} \bigg)\nonumber\\
&+& s_a |\Psi|^2 \bigg(|\Phi_x^a|^2 + |\Phi_y^a|^2\bigg) + s_b |\Psi|^2 \bigg(|\Phi_x^b|^2 + |\Phi_y^b|^2\bigg) \bigg],
\eeq
\end{widetext}
where $r_a=a(T-T_{c,0}^{a})$ and $r_b=b(T-T_{c,0}^{b})$ with $a,b>0$ and the bare transition temperatures,  $T_{c,0}^{b}>T_{c,0}^{a}$. Note that we have already utilized the emergent symmetry of the linearized hot-spot theory to equate the transition temperatures for SC and CDW-b, and also equated the coefficients of $|\Psi|^4$ and $|\Phi_x^b|^4,~|\Phi_y^b|^4$. In the presence of terms that break this symmetry, $T_{c,0}^{b}<T_{c,0}^{\tn{SC}}$; the exact details are beyond the scope of this work.

It is important to note that the above action is invariant under all the underlying symmetries (including under rotations, e.g.  ${\cal{R}}_{\pi/2}: \Phi_x^a\rightarrow\Phi_y^a; \Phi_y^a\rightarrow\Phi_x^{a*}; \Phi_x^b\rightarrow \Phi_y^{b*}; \Phi_y^b\rightarrow \Phi_x^b$). The terms in the third and fourth lines arise naturally due to the existence of two types of CDW correlations in the system with wavevectors that satisfy the following geometric constraints: $\Q^b_x=\Q^a_x+\Q^a_y$ and $\Q^b_y=\Q^a_x-\Q^a_y$. Some of these coefficients were computed for the full fermi-surface in Ref.\cite{norman14}. Note the absence of a term of the form $ |\Phi_x^b|^2|\Phi_y^b|^2$ above, which is allowed by symmetry but missing due to lack of available phase-space for this kind of a scattering process. The terms that are of particular interest to us appear in the last line ($\sim s_a, s_b$), as will become clear in the next section.

Let us now present the results for the different coefficients that appear above. 

\section{Results} 

We start by presenting the results for the linearized theory (i.e. set $\kappa=0$) in the vicinity of the hot-spots. 

\subsection{Linearized hot-spot theory}
\label{reshs}
In this section, we shall list the expressions for the coefficients in terms of the loop integrals. The details of the computation are presented in appendix \ref{int}. At the outset, we note the regime that we are working in here --- we assume that $T\ll v_x\Lambda\ll v_y\Lambda$, i.e. the temperature is much lower than any other ultraviolet energy scale in the problem and futhermore, the regions in the vicinity of the hot-spots (and in the antinodal regions) are almost nested. In the next section, we will show that the fermi-surface curvature, $\kappa$, introduces another temperature scale in the problem above which our analysis remains valid.

We start with $u_a$, representing the coefficient of the $|\Phi_x^a|^4, |\Phi_y^a|^4$ term (fig.\ref{fd}a). Evaluating the contributions arising from both the patches, we obtain,
\beq
u_a&=&-(I_{1616}+I_{2525}+I_{3434}+I_{7878})\\
&=&-(I_{1212}+I_{3838}+I_{4747}+I_{5656}),~\tn{where}
\eeq
it is straightforward to see that $I_{1616}=I_{2525}=I_{3838}=I_{4747}$, and $I_{1212}=I_{3434}=I_{5656}=I_{7878}$.
The loop integrals are given by,
\beq
I_{1212}&=&-\frac{1}{2}\int_\k G_1^2~G_2^2\approx-\frac{1}{16\pi^2v_x^2v_y\Lambda}\frac{1}{(1+(\frac{\pi T}{v_x\Lambda})^2)},\\
I_{2525}&=&-\frac{1}{2}\int_\k G_2^2~G_5^2=0,
\label{Iua}
\eeq
where we use the notation $\int_\k\equiv T\sum_m\int~dk_x~dk_y/(2\pi)^2$ and all the internal Green's functions carry the same argument: $(i\omega_m,\k)$, with $\omega_m=(2m+1)\pi T$. Note that in the limit of $T\ll v_x\Lambda$, $I_{1212}\rightarrow-1/(16\pi^2v_x^2v_y\Lambda)$, i.e. it is non-singular and approaches a constant independent of temperature. 

The competition term between $\Phi_x^a$ and $\Phi_y^a$, described by $w_a$ (also, fig.\ref{fd} a), is given by,
\beq
w_a&=&-(2{\rm{S}}+{\rm{V}}),~\tn{where}\\
{\rm{S}}&=&I_{2565}+I_{5212}+I_{2161}+I_{1656},\\
{\rm{V}}&=&4I_{1256}.
\eeq
In the above, ${\rm{S}},~{\rm{V}}$ represent the self-energy and vertex-correction type diagrams. Furthermore, it is straightforward to see that $I_{2565}=I_{5212}=I_{2161}=I_{1656}$. The explicit expressions are given by,
\beq
I_{2565}&=&-\int_\k G_2~G_5^2~G_6 \approx -\frac{1}{8\pi^2v_x^2v_y\Lambda}\log\bigg(\frac{v_x\Lambda}{\pi T} \bigg),\\
I_{1256}&=& - \int_\k G_1G_2G_5G_6 = -\frac{1}{32v_xv_yT},
\eeq 
where the second integral has been evaluated in the limit $v_x\Lambda\rightarrow\infty$. 
Therefore, we see that the most singular contribution to $w_a$ comes from $I_{1256}$ and is $\sim 1/T$. This has interesting consequences, as will be discussed at the end of this section, and has also been pointed out by a recent work \cite{YWAC14}.

Similarly, the contributions to the $|\Phi^b|^4$ terms arise from (fig.\ref{fd} b),
\beq
u_b&=&-(J_{2626}+J_{3737})=-(J_{1515}+J_{4848}),~\tn{where}
\eeq
due to the underlying symmetries, all the diagrams turn out to be equal, i.e. $J_{1515}=J_{2626}=J_{3737}=J_{4848}$.
The integral evaluates to,
\beq
J_{1515}=-\frac{1}{2}\int_\k G_1^2~G_5^2=-\frac{7\zeta(3)}{32\pi^4}\frac{\Lambda}{v_yT^2},
\eeq
where $\zeta(n)$ is the Riemann-zeta function. The singularity here is much stronger than what we encountered before in the case of the $|\Phi^a|^4$ terms. However, the $T^{-2}$ behavior is not at all surprising--- recall that there is a perfect SU(2)  symmetry between CDW-b and SC within our linearized theory and the coefficient of $|\Phi^b|^4$ term should therefore be identical to that of $|\Psi|^4$, which is known to be of the same $T^{-2}$ form. In the presence of a finite curvature, this symmetry will be broken below some temperature scale set by $\kappa$, as we shall see in the next section. 

Let us now shift our focus to terms that describe the competition between the different components of the charge-orders, $\Phi^a$ and $\Phi^b$. There are two types of four-point functions between these orders, denoted $u_{ab}$ and $\ol{u}_{ab}$. Let us focus on $u_{ab}$ first (fig.\ref{fd}c). If we focus only on the coefficient of, let us say, the $ |\Phi_x^a|^2|\Phi_x^b|^2$ term (the overall form in which the different order-parameters appear is strongly constrained by various  symmetries), we get, 
\beq
u_{ab}&=&-2(K_{1626}+K_{4373}),
\eeq
where $K_{4373}=K_{6515}$ by symmetry under ${\cal{R}}_{\pi/2}$ (and the latter appears in the coefficient of $|\Phi_y^a|^2|\Phi_y^b|^2$). Evaluating these loop integrals gives,
\beq
K_{1626}&=&-\int_{\k}G_1~G_2~G_6^2 = -\frac{1}{64v_xv_yT}, \\
K_{6515}&=&-\int_{\k}G_1~G_5^2~G_6 = -\frac{1}{64v_xv_yT}.
\eeq
Similarly, while evaluating $\ol{u}_{ab}$ (fig.\ref{fd}d), if we focus only on the coefficient of $ (\Phi^a_x)^2\Phi^{b*}_x\Phi^{b*}_y$ (the overall form of the terms is once again constrained by symmetry),
\beq
\ol{u}_{ab}=-(L_{2516}+L_{3784}),
\eeq
where $L_{3784}=L_{2516}$ and the explicit form is given by,
\beq
L_{2516}=-\int_\k G_1G_2G_5G_6 = -\frac{1}{32v_xv_yT}.
\eeq
It is interesting to note that the leading singularities in all of the above diagrams (with the exception of $u_a,~ u_b$) is of the form $\sim 1/T$. This is something that we can understand by applying  standard power-counting arguments. In $(2+1)-$dimensions for such 4-point functions, the singular structure in the IR (with a cutoff, $k_0\sim T$) will be obtained as $\int d^3k/ k^4\sim 1/k_0$, where $k\equiv(i\omega,\k)$. However, there are obviously exceptions to this naive argument, which arise due to the interesting pole structure of the propagators involved in the different diagrams.

We now move over to the terms that actually describe the competition between CDW and SC---these will be responsible for some of the interesting results to come out of our analysis. We start by evaluating the diagrams contributing to $s_a$, which describes competition between $\Phi^a$ and $\Psi$ (fig.\ref{fd} e, f),
\beq
s_a&=&2\ol{S}+\ol{V},~\tn{where}\\
\ol{S}&=&M_{2515}+M_{6151}+M_{7848}+M_{4373},\nonumber\\
&=&M_{1262}+M_{6515}+M_{8373}+M_{4737},\\
\ol{V}&=&N_{2615}+N_{8437}=N_{2651}+N_{8473}.
\eeq
In the above, $\ol{S}$ and $\ol{V}$ represent the self-energy and vertex correction contributions respectively. We have written the coefficients of both $|\Phi_x^a|^2|\Psi|^2$ and $|\Phi_y^a|^2|\Psi|^2$ above, which are of course equal. Moreover, some of the symmetry related diagrams are individually equal as well, such as $M_{1262}=M_{7848}=M_{6515}=M_{4373}$ and $M_{2515}=M_{8373}=M_{4737}=M_{6151}$. Similarly, $N_{2651}=N_{2615}=N_{8437}=N_{8473}$. The explicit expressions for the (distinct) diagrams are given by,
\beq
M_{2515}&=& -\int_\k G_2~G_5^2~G_1' = \frac{1}{64v_xv_yT} \\
M_{1262}&=& -\int_\k G_1~G_2^2~G_6' = \frac{1}{64v_xv_yT},\\
N_{2651}&=& -\int_\k G_1G_2G_5'G_6' = -\frac{1}{32v_xv_yT} .\nonumber\\
\eeq

The ``primed" Green's functions have arguments $(-i\omega_m,-\k)$. Once again, the leading singularity is of the form $1/T$.

Finally, the coefficent $s_b$, which describes the competition between $\Phi^b$ and $\Psi$ (figs.\ref{fd} g and h) is given by,
\beq
s_b&=&2(P_{2626}+P_{3737})+(Q_{2626}+Q_{3737})\\
&=&2(P_{1515}+P_{4848})+(Q_{1515}+Q_{4848}),~\tn{where}
\eeq
all the self-energy type diagrams, $P_{1515}=P_{2626}=P_{3737}=P_{4848}$, are equal and the vertex-correction type diagrams are equal and opposite in sign to the self-energy type ones, i.e. $Q_{1515}=Q_{2626}=Q_{3737}=Q_{4848}=-P_{1515}$.
We therefore only need to evaluate one such integral---the corresponding expression is given by,
\beq
P_{2626}&=&- \int_\k G_2~G_6^2~G_2' =\frac{7\zeta(3)}{16\pi^4}\frac{\Lambda}{v_yT^2}.
\eeq
The $1/T^2$ behavior is to be expected by the same reasoning that was presented earlier --- the coefficients of $|\Psi|^4,~|\Phi^b|^4$ and $|\Phi^b|^2|\Psi|^2$ should have an identical (singular-) structure arising from the emergent SU(2) symmetry. 
\begin{figure}
\begin{center}
\includegraphics[width=0.4\columnwidth]{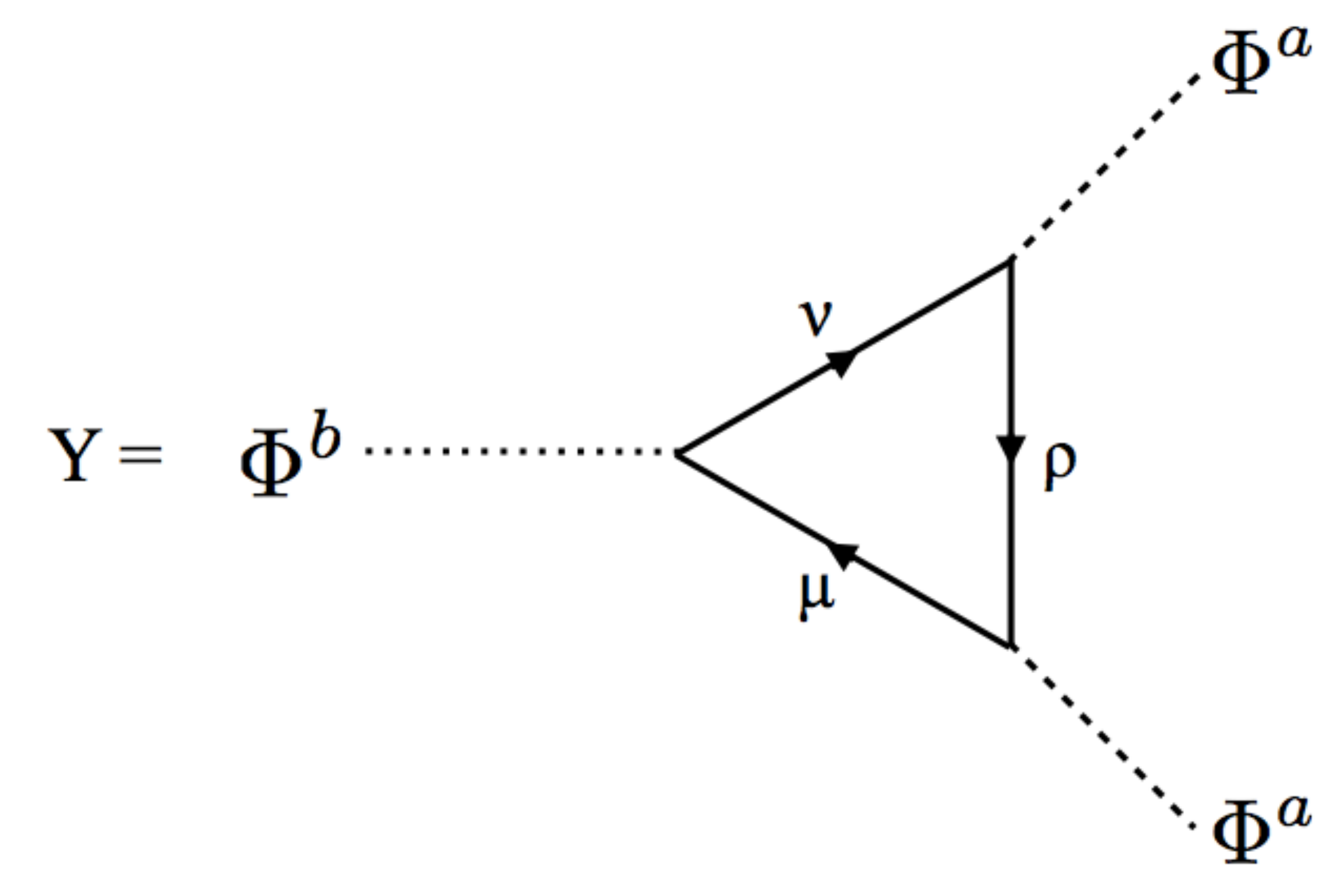}
\end{center}
\caption{Feynman diagram representing the 3-point function, $Y_{\mu\nu\rho}$, between $\Phi^a$ (dashed lines) and $\Phi^b$ (dotted line). The solid internal lines carry different hot-spot indices ($\mu,\nu,\rho$).}
\label{fdt}
\end{figure}

Now that we have evaluated all the four-point functions allowed by symmetry, let us also evaluate the three-point functions between $\Phi^a, \Phi^b$ that contribute to $t_{ab}$ (fig.\ref{fdt}). Once again, we remind the reader that such a term is allowed because of purely geometric reasons associated with the wavevectors of the various CDWs: $\Q^b_x=\Q^a_x+\Q^a_y$ and $\Q^b_y=\Q^a_x-\Q^a_y$. If we focus on the coefficient of the $\sim\Phi_x^a\Phi_y^a\Phi_x^{b*}$ term (and the rest just follows by symmetry), we get,
\beq
t_{ab}=(Y_{261}+Y_{265}+Y_{374}+Y_{378}), \tn{where}
\eeq
it is easy to see that $Y_{265}=Y_{261}=Y_{374}=Y_{378}$. If we choose to evaluate just one of these,
\beq
Y_{261}=-\int_\k G_1~G_2~G_6=0.
\eeq
It is very interesting to see that within our linearized theory, the integral evaluates exactly to $0$. However, in the presence of a finite curvature, this term assumes a non-zero value, as will be shown in the next section.

Assembling all the expressions that we have computed above, the leading (singular-) behavior of the coefficients in the effective action, $S_{\tn{eff}}[\Phi^a,\Phi^b,\Psi]$ are given by,
\beq
u_a&=&\frac{1}{8\pi^2v_x^2v_y\Lambda},~w_a=\frac{1}{8v_xv_yT},~u_b=\frac{7\zeta(3)\Lambda}{16\pi^4v_yT^2},\\
u_{ab}&=&\frac{1}{16v_xv_yT},~ \ol{u}_{ab}=\frac{1}{16v_xv_yT} ,~t_{ab}=0 \\
 s_a&=&\frac{1}{16v_xv_yT},~s_b=\frac{7\zeta(3)\Lambda}{8\pi^4v_yT^2}.
\eeq

At this point, it is worth pointing out some of the interesting features associated with the above terms. First of all, notice that depending on the nature of the term, we have obtained two different types of singularities---there are terms that go as $1/T$, and others that go as $1/T^2$--- in addition to the non-singular term. This has two interesting consequences. In the presence of only the terms involving CDW-a, the competition between the $x,y-$ components, $w_a$, far exceeds $u_a$, i.e. $w_a/u_a\sim \Lambda v_x/T\gg1$. The implication is that at low enough temperatures, CDW-a would necessarily have a tendency to form stripe-like, instead of checkerboard, order \cite{YWAC14} which would spontaneously break the underlying $C_4$ symmetry of the lattice. On the other hand, with only CDW-b order, due to the absence of any competition between its $x,y-$ components, there won't be any tendency to break the $C_4$ symmetry. There is indication for the CDW being unidirectional and stripe-like in the absence of a magnetic field in various experiments.

Let us now discuss an important feature of our analysis, involving the $s_a, s_b$ terms that describe the competition of the different CDWs with SC. We find that $s_b/s_a\sim \Lambda v_x/T\gg1$, implying that at low enough temperatures, SC competes with CDW-b much more strongly than with CDW-a. However, this is not surprising for the following reason: In our linearized hot-spot theory, there is no fundamental difference between CDW-b and SC due to the SU(2) symmetry. In fact both of these orders are strongly coupled to each other and compete for density of states on the fermi-surface in the vicinity of the same hot-spots. Therefore, it is natural for them to compete with each other more strongly. The same is not true about CDW-a and SC, which compete for density of states along different portions of the Fermi-surface. Note that CDW-a and CDW-b also compete mutually, so suppressing one naturally makes it favorable for the other one to emerge.  

\begin{figure}
\begin{center}
\includegraphics[width=0.5\columnwidth]{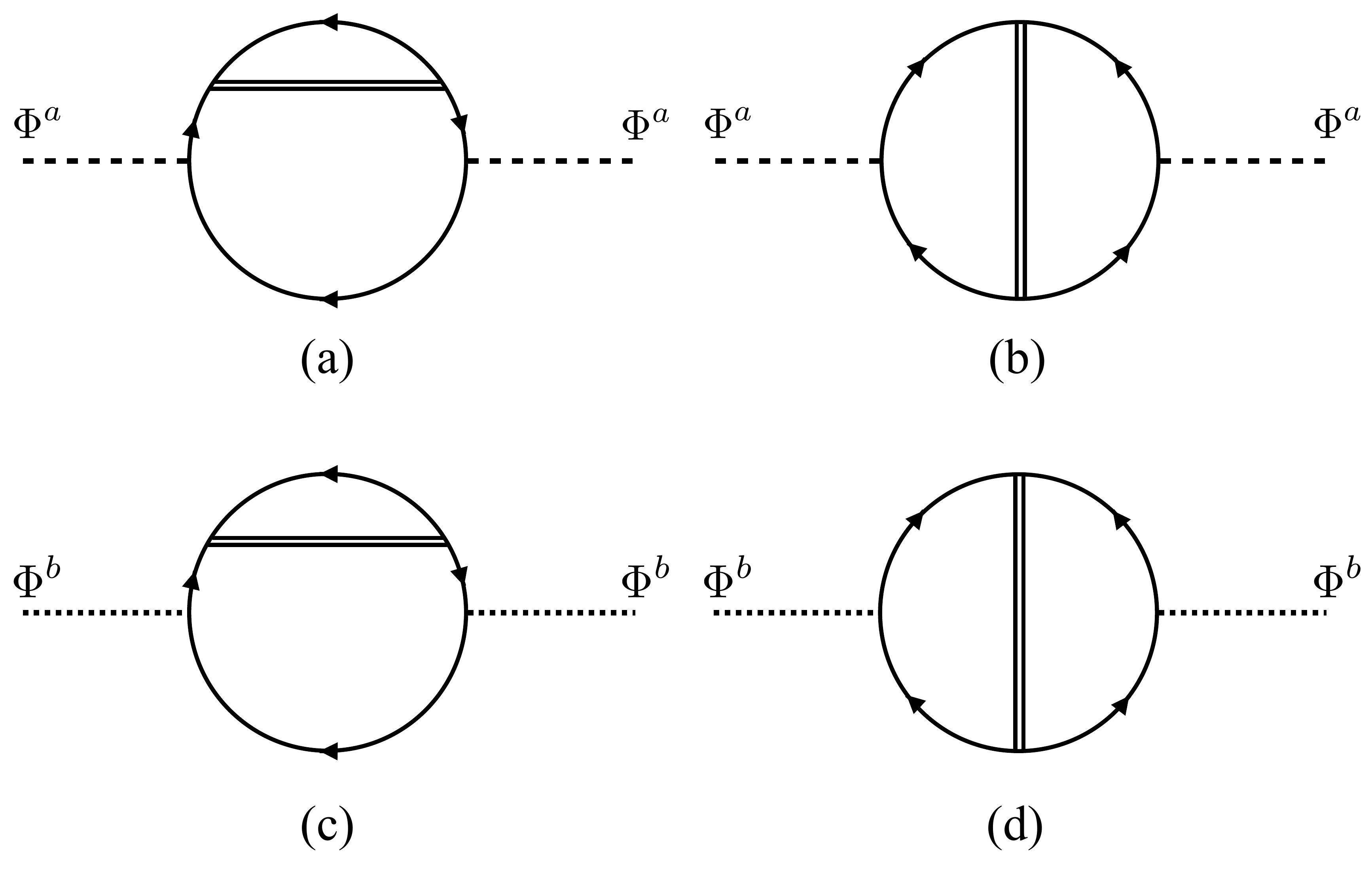}
\end{center}
\caption{Contracting the $\Psi-$ fields (solid-double lines) in (a) $M_{\mu\nu\rho\delta}$, (b) $N_{\mu\nu\rho\delta}$ renormalize the $|\Phi^a|^2$ term, and, (c) $P_{\mu\nu\rho\delta}$, (d) $Q_{\mu\nu\rho\delta}$ renormalize the $|\Phi^b|^2$ term.    }
\label{ren}
\end{figure}

To summarize, the results of this section indicate that at very high temperatures, there is an almost perfect symmetry between CDW-b and SC (in fact, this symmetry persists even in the presence of a finite curvature, as we shall show in the next section) which makes it unfavorable for CDW-a to appear in the scene. However, as a function of decreasing temperature, as the strength of superconducting fluctuations increase, the CDW-b fluctuations are preferentially suppressed compared to CDW-a, which could possibly allow CDW-a to emerge. The exact crossovers, if any, are beyond the scope of this work. The net effect of the SC fluctuations is to effectively renormalize the quadratic terms in the action for $\Phi^a$ and $\Phi^b$, as shown in fig.\ref{ren}. At low temperatures, the coefficient of $|\Phi^b|^2$ is renormalized more strongly compared to the coefficient of $|\Phi^a|^2$. In order to extract an estimate of the relative renormalizations, let us compute $\D r_a$ and $\D r_b$ for CDW-a,b due to purely thermal and Gaussian fluctuations associated with superconductivity, $\langle |\Psi_\q|^2\rangle\sim (\q^2+\sigma^2)^{-1}$. This approximation is valid sufficiently far away from the superconducting $T_c$. The renormalizations are then given by,
\beq
\D r_a &\approx& s_a\int \frac{d^2q}{(2\pi)^2}\frac{1}{q^2+\sigma^2} \approx -\frac{s_a}{2\pi}\log\bigg(\frac{\sigma}{\Lambda} \bigg) =  -\frac{1}{32\pi v_xv_y T}\log\bigg(\frac{\sigma}{\Lambda} \bigg),\\
\D r_b &\approx& s_b\int \frac{d^2q}{(2\pi)^2}\frac{1}{q^2+\sigma^2} \approx -\frac{s_b}{2\pi}\log\bigg(\frac{\sigma}{\Lambda} \bigg) = -\frac{7\zeta(3)\Lambda}{16\pi^5v_yT^2} \log\bigg(\frac{\sigma}{\Lambda} \bigg).
\eeq
Hence, the relative renormalizations $\D r_b/\D r_a\sim \Lambda v_x/T \gg 1$, which indicates that within our simplified  theory, there exists an intermediate range of temperatures where the transition temperature for CDW-b gets suppressed more than the one for CDW-a. 

The only caveat in the calculations presented in this section is that all of them were performed in the limit of $\kappa\rightarrow0$. Of course, in the actual problem, the curvature is finite (albeit small, in the anti-nodal region). Therefore, we revisit the whole problem with a non-zero curvature in the next section and analyze the consequences numerically. 

We would like to remind the reader, that at the level of approximation that we have used in this section, the CDW-a state is degenerate with the corresponding PDW state and the two compete strongly with each other. However, in the presence of a curvature or other features that break the SU(2) symmetry, the PDW state is destroyed completely, at least within the weak-coupling picture (unlike the other states, as we saw in fig.\ref{eigfig}). Moreover, as we shall see in the next section, the competition effects between CDW-a,b and SC discussed above survive even in the presence of curvature as long as the temperature is higher than a scale set by $\kappa$.
  
\subsection{Effect of fermi-surface curvature}
\label{rescu}
In the previous section we ignored the effect of the fermi-surface curvature completely and analyzed the linearized theory in the vicinity of the hot-spots. Most of the graphs that we evaluated were singular in the limit of $T\rightarrow0$ (though we intend to apply our results to the metallic state at strictly $T>0$). The question we would now like to address is to what extent do these results remain valid and what is the regime of validity, in the presence of a finite curvature. In fact, it is possible that the curvature sets a temperature scale above which our analytical results for the linearized theory continue to hold. We shall denote all the previously evaluated integrals by $\tilde{I}, \tilde{J},..., \tilde{Q}, \tilde{Y}$, to distinguish them from the symbols used earlier. This time around, we shall perform the Matsubara sums first and then evaluate the momentum-integrals numerically as a function of temperature for various fixed values of curvature, $\kappa$, and $\alpha=v_y/v_x$. 

The aim of this computation is two-fold. First of all, in the limit of $\kappa\rightarrow0$, we should recover the temperature dependencies of the different terms obtained earlier. Secondly, we would like to have an approximate estimate of the functional form of $T_0(\kappa,\alpha)$, the temperature scale above which our analytical results for the $\kappa=0$ problem continue to hold, assuming such a scale exists. 

In the presence of a finite curvature, the modified dispersions are now given by,
\beq
\e_1(\k)&=&v_xk_x+v_yk_y+\kappa(k_x^2+k_y^2),\\
\e_2(\k)&=&v_xk_x-v_yk_y+\kappa(k_x^2+k_y^2),
\eeq
with $\e_5(\k)=\e_1(-\k)$, $\e_6(\k)=\e_2(-\k)$ and so on.

We only present the main findings of this analysis in the present section. The technical details of the computations alongwith plots of the numerical results are provided in Appendix \ref{intcurv}.

Let us start by analyzing the temperature dependence of diagrams ($\tilde{I}_{1212}$) contributing to $u_a$. We found that (i) at a fixed value of $\alpha$, but for $\kappa\neq0$, $\tilde{I}_{1212}$ goes to a constant value in the limit of $T\rightarrow0$. Moreover, in the limit of $\kappa\rightarrow0$, this constant is nearly identical to what we had computed earlier for $I_{1212}$ (eqn.\ref{Iua}). (ii) On the other hand, for $T\rightarrow v_x\Lambda$, there is a power-law fall off going as $\sim 1/T^2$ for all considered values of $\kappa$, which again matches with our analytical result for $I_{1212}$ (eqn.\ref{Iua}). (iii) Finally, $\tilde{I}_{1212}$ scales as $\sim 1/\alpha$, which was apparent from the perfect scaling collapse that we observed for $\alpha\tilde{I}_{1212}$ (not shown). This agrees with our analytical results from earlier, even though they were computed with $\kappa=0$. The numerically evaluated results for $\tilde{I}_{1212}$ are shown in fig.\ref{curvI}(a).

We next computed the leading diagrams contributing to $w_a$, giving rise to competition between $x-$ and $y-$ components of CDW-a. We came across the following interesting results: (i) irrespective of the value of the curvature, the diagrams all asymptote to a $1/T$ behavior at low temperatures (upto $T/v_x\Lambda \sim 10^{-4}$). (ii) There were however deviations at higher temperatures. (iii) Finally, this particular diagram also scales as $1/\alpha$ (not shown). The plots as a function of temperature are shown in fig.\ref{curvI}(b). 

It is especially interesting to see that even in the presence of a small curvature, $w_a$ continues to scale as $1/T$ down to very low temperatures. However, as a function of decreasing temperature, there will be a preemptive instability to superconductivity, which in turn will cut-off the $1/T$ behavior to $1/\Delta$, where $\Delta$ is the superconducting gap.

Let us now revisit the terms that turned out to be the most singular in our earlier analysis, which includes $u_b$ ($J_{1515}$) and $s_b$ ($P_{2626},~Q_{2626}$), and went as $\sim 1/T^2$.  As a reminder, in the linearized theory, we obtained $2J_{1515}=-P_{2626}=Q_{2626}$. However, a finite curvature breaks this symmetry. We first focussed on the temperature-dependence of these diagrams at a fixed $\alpha$ but different values of $\kappa$ and noticed the following common features: (i) The limit of $\kappa\rightarrow0$ computation agrees perfectly with the analytical computation from the previous section. (ii) With an increasing $\kappa$, we note that the results for the different computations (i.e. with and without $\kappa$) only agree with each other above a characteristic temperature, $T_0=\rm{C}\kappa$, with $\rm{C}$ different for each diagram (This is determined by noting the temperature, $T_0$, at which the deviation starts; these are marked by the dotted vertical lines in fig.\ref{curv1t2}). To investigate whether ${\rm{C}}$ is $\alpha-$dependent, we computed the same diagrams as a function of temperature at a fixed $\kappa\neq0$ and $v_x$, but for different values of $\alpha$. Remarkably, the value of $T_0$ remains unaffected by changing $\alpha$, which shows that ${\rm{C}}$ is independent of $\alpha$. (iii)  We also observed them to scale as $\sim1/\alpha$, just like in all the previous cases. Therefore, to summarize, $u_b$ and $s_b$ continue to behave as $1/T^2$ above a temperature scale that is set by curvature, $T_0\sim\kappa$. Therefore, at high enough temperatures compared to this scale, the degeneracy between CDW-b and SC is maintained and the competition is sufficiently weak that it is unlikely that CDW-a will be a preferred state, based on our arguments from the previous section. It is however important to remind ourselves that the coefficient of $|\Psi|^4$, which also goes as $\sim 1/T^2$, survives even in the presence of a finite curvature but is eventually cut-off by $\Delta$. The numerically evaluated results are shown as a function of temperature in fig.\ref{curv1t2} (a)-(f).

Finally, we also evaluated the term responsible for competition between SC and CDW-a. Recall that in the linearized theory, $2M_{2515}=-N_{2651}=-N_{2615}$ and where the leading singularities were all $\sim 1/T$. However, in the presence of a finite $\kappa$ these degeneracies are lifted. However, all the diagrams continue to behave as $\sim 1/T$ down to temperatures of $T/v_x\Lambda\sim 10^{-4}$, even in the presence of a reasonably large $\kappa$. However, as a function of decreasing temperature, the system will go superconducting thereby cuting-off the singularity. We have also checked that just like all the other diagrams considered so far, these diagrams also scale as $\sim 1/\alpha$. The results are shown in fig.\ref{curvMN} (a)-(c).

Towards the end of section \ref{reshs}, we saw that the three-point functions between CDW-a and b turned out to be identically zero. However, when we evaluated the same diagrams in the presence of a finite curvature, they turned out to be non-zero. In fact, based on the numerically evaluated results, we were able to guess an analytical functional form for $\tilde{Y}_{261}$, which is as follows,
\beq
{\mathcal{Y}}(T,\kappa,\alpha)\approx\frac{\kappa\Lambda}{\pi^2\alpha v_x}\frac{1}{\pi^2T^2+v_x^2\Lambda^2}.
\eeq
Note that it indeed reproduces the $\kappa\rightarrow0$ limit correctly and approaches a constant in the limit of $T\rightarrow0$ otherwise. The numerically evaluated results and a comparison with ${\mathcal{Y}}$ is shown in fig.\ref{y261}(a), (b). 

To conclude this section, we found that at sufficiently high temperatures, our results from the previous section continue to hold even in the presence of a finite, but small, curvature. The terms that we found to be most singular in our earlier computation ($\sim 1/T^2$), continue to have the same form as long as $T>T_0\sim {\rm{C}}\kappa$. On the other hand, the terms that went as $\sim 1/T$, continue to be so down to very low temperatures compared to the scales set by the fermi-velocities. However, as a function of decreasing temperature, these singularities are cut-off eventually by the preemptive instability to superconductivity. It is therefore safe to conclude that our computations in section \ref{reshs} are applicable in the window $\tn{max}\{T_c, T_0\}<T<v_x\Lambda\ll v_y\Lambda$. 

\section{Discussion}
\label{dis}
Over the past few years, we have learnt a great deal about the nature of the various symmetry-broken states that arise in the pseudogap regime of the underdoped cuprates. This has largely been possible due to the enormous number of remarkable experiments performed on these materials. Most of these experiments point toward the existence of a fluctuating and short-ranged charge-density wave in a metallic state; the onset of the CDW happens below a characteristic scale $T_{\tn{cdw}}\lesssim T^*$, as deduced from X-ray scattering measurements \cite{LHSS14}. There is a considerable amount of evidence suggesting that the CDW competes with superconductivity. It is therefore essential to understand the true nature of the CDW and its relation to superconductivity, as this might be the key to gaining a complete understanding of the pseudogap phase out of which both orders emerge \cite{AADCSS14}. 

Theoretically, we have now started to realize that the cuprates are a model system where the Fermi surface geometry, the strong interactions between the constituent electrons and the quasi-two dimensional structure conspire to give rise to some remarkable consequences. One of these is the universal feature that in the presence of strong antiferromagnetic interactions, superconductivity and charge-order are tied to each other; this has been highlighted by the observation \cite{comin2,SSJSD14} of the predicted \cite{MMSS10,SSRP13}
$d$-wave form factor of the CDW. While SC and CDW necessarily arise as dual instabilites of the same normal state, they also compete with each other. One of the puzzling features, on the theoretical side, has been the discrepancy between the wavevector of the CDW seen experimentally and the one obtained from the leading CDW instabilities in various models. The primary purpose of this paper has been to address one interesting ingredient that could be partly responsible for resolving this discrepancy over at least an intermediate window of temperature. The primary motivation for invoking the effect of $d-$wave superconducting fluctuations was to suppress density of states in the antinodal regions. 

In this work, we studied the interplay of fluctuating charge-order and superconductivity. Our starting point was the t-J model (without Gutzwiller projection) for a metal interacting via short range antiferromagnetic exchange interactions, where the various instabilities at the Hartree-Fock level are to SC and CDWs with different sets of wavevectors (a t-J-V model with an infinite on-site Hubbard U also leads to similar instabilites, in addition to a staggered flux state \cite{AASS14}). The leading CDW-b state was found to have a wavevector of the form $\pm (Q_0,\pm Q_0)$, while there was a sub-leading instability to the CDW-a with wavevectors $(\pm Q_0,0)$ and $(0,\pm Q_0)$. It is the latter that is closely related to the state seen in the experiments. In order to study the minimal model with all the necessary ingredients, we then considered the theory of a metal with pairing fluctuations and both types of CDW correlations and computed the effective Ginzburg-Landau (GL) theory upto the quartic order in in the low-energy limit. 

We obtained a number of interesting results for the temperature dependencies of the coefficients in the GL theory. In particular, one of the central results of this paper is the nature of the competition between CDW-a and b with SC. We observed that SC competes with CDW-b much more strongly than with CDW-a at low enough, but non-zero  temperatures. In the low-energy limit, we attributed this to the emergent SU(2) symmetry between SC and CDW-b, which really doesn't distinguish between the two different phases, and the absence of a gap in the spectrum at the antinodes. At low temperatures, we presented hints that the SC fluctuations might make it more favorable for CDW-a to arise and CDW-b to be suppressed preferentially. In fact, we showed that even in the presence of a finite fermi-surface curvature, the results for the mutual competition between CDW-a,b and SC continue to hold above a temperature scale that is set by the curvature $(T_0\sim\kappa)$. However at the same time, it is important to note that in the $\kappa\rightarrow0$ limit, CDW-a is related by SU(2) symmetry to the PDW state with the same wavevectors and these two orders would therefore compete strongly with each other. However when the SU(2) symmetry is broken explicitly by a finite fermi-surface curvature (or by a nearest-neighbor Coulomb repulsion term), the fragile PDW state disappears completely, as witnessed in our HF computation for the t-J model. It would be interesting to explore the interplay between SC, CDW and PDW orders beyond weak-coupling, starting from a microscopic model in the near future.

Finally, if it is indeed the superconducting flucutations that are responsible for giving rise to the experimentally observed wavevector, then it is possible that the CDW-b state would show up in experiments if one were to suppress these SC fluctuations completely. Furthermore, the CDW-b order should have tendency to form checkerboard order, unlike CDW-a which has a tendency to form stripe-like order \cite{PHSR13, YWAC14}. This is a direction worth exploring in STM experiments at really high magnetic fields, for instance, and repeating  phase-resolved analysis similar to what has been carried out recently to look for signatures of the CDW-a state \cite{SSJSD14}. However, if the CDW-b state continues to be absent, then it is likely that there are other factors at play here in addition to the SC fluctuations. One such factor has already been considered recently, which arises from strong correlation effects due to Coulomb repulsion \cite{AASS14}.

\acknowledgements
We thank A. Allais, J. Bauer, E. Berg, F.~Lalibert\'e, J. Lee, M. Metlitski, M.~Norman, M. Punk, J. Sau , L.~Taillefer, Y.~Wang and especially A. Chubukov for helpful discussions. D.C. thanks the Weizmann Institute for hospitality, where a part of this work was done. The research was supported by the U.S.\ National Science Foundation under grant DMR-1103860, and by the Templeton Foundation. This research was also supported in part by Perimeter Institute for Theoretical Physics; research at Perimeter Institute is supported by the Government of Canada through Industry Canada and by the Province of Ontario through the Ministry of Research and Innovation.

\appendix
\begin{widetext}
\section{Feynman diagrams for linearized hot-spot theory}
\label{int}
In this appendix, we provide details of the calculations for some of the loop-integrals evaluated earlier. The momentum integrals will all be done with a cutoff $\Lambda$, since we only want to restrict ourselves to the neighborhood of the hot-spots. We start with the diagrams that contribute to $u_a$,
\beq
I_{2525}&=&-\frac{T}{2}\sum_m\int_{|\k|<\Lambda}\frac{1}{(i\omega_m-(v_xk_x-v_yk_y))^2}\frac{1}{(i\omega_m-(-v_xk_x-v_yk_y))^2},\\
I_{1212}&=&-\frac{T}{2}\sum_m\int_{|\k|<\Lambda}\frac{1}{(i\omega_m-(v_xk_x+v_yk_y))^2}\frac{1}{(i\omega_m-(v_xk_x-v_yk_y))^2}.
\eeq
It is useful to change the coordinates to $x=v_xk_x$, $y=v_yk_y$ so that $\Lambda_{x,y}=\Lambda v_{x,y}$. We are in the regime where $\Lambda_y\gg\Lambda_x\gg T$.  The above integrals then become,
\beq
I_{2525}&=&-\frac{T}{8\pi^2v_xv_y}\sum_m\int_{-\Lambda_x}^{\Lambda_x}~dx~\int_{-\Lambda_y}^{\Lambda_y}~dy~\frac{1}{(i\omega_m-x+y)^2}\frac{1}{(i\omega_m+x+y)^2},\\
I_{1212}&=&-\frac{T}{8\pi^2v_xv_y}\sum_m\int_{-\Lambda_x}^{\Lambda_x}~dx~\int_{-\Lambda_y}^{\Lambda_y}~dy~\frac{1}{(i\omega_m-x-y)^2}\frac{1}{(i\omega_m-x+y)^2}.
\eeq
We shall always evaluate the $k_y$ integral first and use $\int_{-\Lambda_y}^{\Lambda_y}dk_y=\int_{-\infty}^\infty dk_y-\int_{|k_y|>\Lambda_y}dk_y={\cal{I}}_1-{\cal{I}}_2$.

The contribution to $I_{2525}$ from ${\cal{I}}_1$, i.e. $I_{2525}\bigg|_1=0$ (poles on same side). From ${\cal{I}}_2$, we get,
\beq
I_{2525}\bigg|_2\approx \frac{T\Lambda_x}{4\pi^2v_xv_y}\sum_m\int_{\Lambda_y}^\infty ~dy~\bigg[\frac{1}{(y+i\omega_m)^4} + \frac{1}{(y-i\omega_m)^4} \bigg]=0
\eeq
Therefore, we have $I_{2525}=0$. For $I_{1212}$, we get,
\beq
I_{1212}\bigg|_1&=&-\frac{iT}{16\pi v_x v_y}\sum_m\tn{sgn}(\omega_m)\int_{-\Lambda_x}^{\Lambda_x}\frac{dx}{(x-i\omega_m)^3}=-\frac{T}{8\pi v_xv_y}\int_{-\Lambda_x}^{\Lambda_x}dx\sum_{m>0}\frac{\omega_m^3-3x^2\omega_m}{(\omega_m^2+x^2)^3}\nonumber\\
&=&-\frac{\Lambda_xT}{4\pi v_xv_y}\sum_{m>0}\frac{\omega_m}{(\omega_m^2+\Lambda_x^2)^2} \approx -\frac{1}{16\pi^2v_xv_y}\frac{\Lambda_x}{\Lambda_x^2+\pi^2T^2}.\\
I_{1212}\bigg|_2&\approx&-\frac{T\Lambda_x}{2\pi^2v_xv_y}\sum_m\int_{\Lambda_y}^\infty~dy~\frac{1}{(y^2+\omega^2)^2}\approx-\frac{1}{16\pi^2v_xv_y}\frac{\Lambda_x}{\Lambda_y^2}.
\eeq

Since $\Lambda_y\gg\Lambda_x$, we can ignore the second contribution and $I_{1212}\approx I_{1212}\bigg|_1$. Note that we have made the approximation, $T\sum_{m>0} F(\omega_m)\approx\int_{\pi T}^\infty \frac{d\omega}{2\pi} F(\omega)$.

Similarly, we have the following contributions to $w_a$, 
\beq
I_{2565}&=&-T\sum_m\int_{|\k|<\Lambda}\frac{1}{(i\omega_m+v_xk_x+v_yk_y)^2}\frac{1}{(i\omega_m-v_xk_x+v_yk_y)}\frac{1}{(i\omega_m+v_xk_x-v_yk_y)},\\
I_{1256}&=&-T\sum_m\int_{|\k|<\Lambda}\frac{1}{(i\omega_m-v_xk_x-v_yk_y)}\frac{1}{(i\omega_m-v_xk_x+v_yk_y)}\frac{1}{(i\omega_m+v_xk_x+v_yk_y)}\frac{1}{(i\omega_m+v_xk_x-v_yk_y)}.\nonumber\\
\eeq
Then,
\beq
I_{2565}&=&-\frac{T}{4\pi^2v_xv_y}\sum_m\int_{-\Lambda_x}^{\Lambda_x}~dx~\int_{-\Lambda_y}^{\Lambda_y}~dy~\frac{1}{(i\omega_m+x+y)^2}\frac{1}{(i\omega_m-x+y)}\frac{1}{(i\omega_m+x-y)},\\
I_{2565}\bigg|_1&=&\frac{T}{16\pi v_xv_y}\sum_m\int_{-\Lambda_x}^{\Lambda_x}~dx~\frac{\tn{sgn}(\omega_m)}{\omega_m}\frac{1}{(x+i\omega_m)^2}\approx-\frac{1}{8\pi^2v_xv_y\Lambda_x}\log\bigg(\frac{\Lambda_x}{\pi T} \bigg),\\
I_{2565}\bigg|_2&\approx&\frac{\Lambda_x T}{2\pi^2v_xv_y}\sum_m\int_{|y|>\Lambda_y}\frac{1}{(i\omega_m+y)^2}\frac{1}{y^2+\omega_m^2}\approx\frac{\Lambda_x}{4\pi^3v_xv_y}\int_{-\infty}^\infty d\omega~\int_{\Lambda_y}^\infty dy~\frac{2(y^2-\omega^2)}{(y^2+\omega^2)^3},\nonumber\\ \\
I_{2565}\bigg|_2&\approx&\frac{1}{16\pi^2v_xv_y}\frac{\Lambda_x}{\Lambda_y^2}.
\eeq

Therefore, we can approximate $I_{2565}\approx I_{2565}\bigg|_1$. Similarly, we have,
\beq
I_{1256}&=&-\frac{T}{4\pi^2v_xv_y}\sum_m\int_{-\Lambda_x}^{\Lambda_x}dx~\int_{-\Lambda_y}^{\Lambda_y}dy~\frac{1}{(i\omega_m-x-y)}\frac{1}{(i\omega_m-x+y)}\frac{1}{(i\omega_m+x+y)}\frac{1}{(i\omega_m+x-y)},\nonumber\\ \\
I_{1256}\bigg|_1&=&-\frac{T}{8\pi v_xv_y}\sum_m\int_{-\Lambda_x}^{\Lambda_x}dx~\frac{\tn{sgn}(\omega_m)}{\omega_m}\frac{1}{\omega_m^2+x^2}.
\eeq
The above integral is convergent in the limit of $\Lambda_x\rightarrow\infty$ (and the singularity comes from small momenta), so that,
\beq
I_{1256}\bigg|_1&=&-\frac{1}{4\pi^2v_xv_yT}\sum_{m>0}\frac{1}{(2m+1)^2}=-\frac{1}{32v_xv_yT}.
\eeq
The other contribution is given by,
\beq
I_{1256}\bigg|_2&=&-\frac{T\Lambda_x}{2\pi^2v_xv_y}\sum_m\int_{|y|>\Lambda_y}dy~\frac{1}{(y^2+\omega_m^2)^2}=-\frac{\Lambda_x}{8\pi^2v_xv_y\Lambda_y^2}.
\eeq
Let us now compute the diagram(s) contributing to $u_b$. They are given by,
\beq
J_{1515}&=&-\frac{T}{2}\sum_m\int_{|\k|<\Lambda}\frac{1}{(i\omega_m-v_xk_x-v_yk_y)^2}\frac{1}{(i\omega_m+v_xk_x+v_yk_y)^2}.
\eeq
This simplifies to,
\beq
J_{1515}&=&-\frac{T}{8\pi^2v_xv_y}\sum_m\int_{-\Lambda_x}^{\Lambda_x}dx \int_{-\Lambda_y}^{\Lambda_y}dy\frac{1}{(i\omega_m-x-y)^2} \frac{1}{(i\omega_m+x+y)^2},\\
J_{1515}\bigg|_1&=&-\frac{T\Lambda_x}{8\pi v_xv_y}\sum_m\frac{\tn{sgn}(\omega_m)}{\omega_m^3},\\
J_{1515}\bigg|_1&=&-\frac{7\zeta(3)}{32\pi^4}\frac{\Lambda}{v_yT^2},
\eeq
The contribution from ${\cal{I}}_2$ turns out to be $J_{1515}\bigg|_2=I_{1212}\bigg|_2$ and can therefore be ignored, compared to $J_{1515}\bigg|_1$.

We now evaluate the contribution to the terms that lead to competition between the different BO, via the 4-point couplings, $u_{ab}$ and $\ol{u}_{ab}$.
\beq
K_{1626}&=&-T\sum_m\int_{|\k|<\Lambda}\frac{1}{(i\omega_m+v_xk_x-v_yk_y)^2}\frac{1}{(i\omega_m-v_xk_x+v_yk_y)}\frac{1}{(i\omega_m-v_xk_x-v_yk_y)},\\
K_{6515}&=&-T\sum_m\int_{|\k|<\Lambda}\frac{1}{(i\omega_m+v_xk_x+v_yk_y)^2}\frac{1}{(i\omega_m-v_xk_x-v_yk_y)}\frac{1}{(i\omega_m+v_xk_x-v_yk_y)},\\
L_{2516}&=&-T\sum_m\int_{|\k|<\Lambda}\frac{1}{(i\omega_m-v_xk_x-v_yk_y)}\frac{1}{(i\omega_m-v_xk_x+v_yk_y)}\frac{1}{(i\omega_m+v_xk_x+v_yk_y)}\frac{1}{(i\omega_m+v_xk_x-v_yk_y)}.\nonumber\\
\eeq
It is straightforward to see that $K_{1626}=-M_{2515}$, $K_{6515}=-M_{1262}$, and $L_{2516}=N_{2651}(=I_{1256})$ which we evaluate in detail below. 

Let us now evaluate the terms contributing to the competition terms, $s_a$ and $s_b$, between CDW and SC. We start with the distinct self-energy type diagrams contributing to $s_a$,
\beq
M_{2515}&=&-T\sum_m\int_{|\k|<\Lambda}\frac{1}{(i\omega_m+v_xk_x+v_yk_y)^2}\frac{1}{(i\omega_m-v_xk_x+v_yk_y)}\frac{1}{(-i\omega_m+v_xk_x+v_yk_y)},\nonumber \\ \\
M_{1262}&=&-T\sum_m\int_{|\k|<\Lambda}\frac{1}{(i\omega_m-v_xk_x+v_yk_y)^2}\frac{1}{(i\omega_m-v_xk_x-v_yk_y)}\frac{1}{(-i\omega_m-v_xk_x+v_yk_y)}.\nonumber\\
\eeq
Transforming to the $x,y-$coordinates, this becomes,
\beq
M_{2515}&=&-\frac{T}{4\pi^2v_xv_y}\sum_m\int_{-\Lambda_x}^{\Lambda_x}~dx~\int_{-\Lambda_x}^{\Lambda_x}~dy~\frac{1}{(i\omega_m+x+y)^2}\frac{1}{(i\omega_m-x+y)}\frac{1}{(-i\omega_m+x+y)},\nonumber\\ \\
M_{1262}&=&-\frac{T}{4\pi^2v_xv_y}\sum_m\int_{-\Lambda_x}^{\Lambda_x}~dx~\int_{-\Lambda_x}^{\Lambda_x}~dy~\frac{1}{(i\omega_m-x+y)^2}\frac{1}{(i\omega_m-x-y)}\frac{1}{(-i\omega_m-x+y)}.\nonumber\\
\eeq
By splitting the integral as earlier, we have from ${\cal{I}}_1$,
\beq
M_{2515}\bigg|_1&=&-\frac{iT}{16\pi v_xv_y}\sum_m\int_{-\Lambda_x}^{\Lambda_x}~dx~\frac{\tn{sgn}(\omega_m)}{\omega_m^2}\frac{1}{x-i\omega_m}=\frac{T}{8\pi v_xv_y}\sum_{m>0}\int_{-\Lambda_x}^{\Lambda_x}~dx~\frac{1}{\omega_m(x^2+\omega_m^2)},\nonumber \\ \\
M_{2515}\bigg|_1&=&\frac{1}{64v_xv_yT},
\eeq
where we have used the fact that $2M_{2515}\bigg|_1=-I_{1256}\bigg|_1$.
Similarly from ${\cal{I}}_2$, we get,
\beq
M_{2515}\bigg|_2&\approx&-\frac{T\Lambda_x}{2\pi^2v_xv_y}\sum_m\int_{|y|>\Lambda_y}dy~\frac{1}{y^2+\omega_m^2}\frac{1}{(y+i\omega_m)^2}\approx-\frac{\Lambda_x}{4\pi^3v_xv_y}\int_{-\infty}^\infty d\omega~\int_{\Lambda_y}^\infty dy~\frac{2(y^2-\omega^2)}{(y^2+\omega^2)^3},\nonumber\\ \\
M_{2515}\bigg|_2&\approx&-\frac{\Lambda_x}{16\pi^2v_xv_y\Lambda_y^2}.
\eeq

For $M_{1262}$,
\beq
M_{1262}\bigg|_1&=&-\frac{iT}{16\pi v_xv_y}\sum_m\int_{-\Lambda_x}^{\Lambda_x}dx~\frac{x-2i\omega_m}{(x-i\omega_m)^2\omega_m^2}\tn{sgn}(\omega_m)=-\frac{iT}{16\pi v_xv_y}\sum_{m>0}\int_{-\Lambda_x}^{\Lambda_x}dx~\frac{4i\omega_m}{(x^2+\omega_m^2)^2},\nonumber\\ \\
M_{1262}\bigg|_1&=&\frac{1}{64 v_xv_yT}.\\
M_{1262}\bigg|_2&\approx&\frac{T\Lambda_x}{2\pi^2v_xv_y}\sum_m\int_{|y|>\Lambda_y}dy~\frac{1}{(y^2+\omega_m^2)^2}=\frac{\Lambda_x}{8\pi^2v_xv_y\Lambda_y^2}.
\eeq
We can ignore $M_{2515}\bigg|_2, M_{1262}\bigg|_2$ compared to $M_{2515}\bigg|_1,M_{1262}\bigg|_1$.

Let us now evaluate the only distinct vertex-correction type diagram contributing to $s_a$,
\beq
N_{2651}&=&-T\sum_m\int_{|\k|<\Lambda}\frac{1}{(i\omega_m-v_xk_x-v_yk_y)}\frac{1}{(i\omega_m-v_xk_x+v_yk_y)}\frac{1}{(-i\omega_m-v_xk_x-v_yk_y)}\frac{1}{(-i\omega_m-v_xk_x+v_yk_y)}.\nonumber\\
\eeq
Upon transforming coordinates, we have
\beq
N_{2651}&=&-\frac{T}{4\pi^2v_xv_y}\sum_m\int_{-\Lambda_x}^{\Lambda_x}dx~\int_{-\Lambda_y}^{\Lambda_y}dy~\frac{1}{(i\omega_m-x-y)}\frac{1}{(i\omega_m-x+y)}\frac{1}{(-i\omega_m-x-y)}\frac{1}{(-i\omega_m-x+y)},\nonumber\\
\eeq
and we immediately see that $N_{2651}=I_{1256}$, as expected.

The diagrams contributing to $s_b$ can also be evaluated analogously as follows:
\beq
P_{2626}&=&-T\sum_m\int_{|\k|<\Lambda}\frac{1}{(i\omega_m-v_xk_x+v_yk_y)}\frac{1}{(i\omega_m+v_xk_x-v_yk_y)^2}\frac{1}{(-i\omega_m+v_xk_x-v_yk_y)}.
\eeq 
However, we immediately notice that $P_{2626}=-2J_{2626}=-2J_{1515}$ (which we have already evaluated above), due to the underlying SU(2) symmetry of the hot-spot theory.

Finally, let us compute the diagram contributing to the three-point function, $Y_{\mu\nu\rho}$. The one we intend to compute is,
\beq
Y_{261}&=&-T\sum_m\int_{|\k|<\Lambda}\frac{1}{(i\omega_m-v_xk_x-v_yk_y)}\frac{1}{(i\omega_m-v_xk_x+v_yk_y)}\frac{1}{(i\omega_m+v_xk_x-v_yk_y)},\\
Y_{261}&=&-\frac{T}{4\pi^2v_xv_y}\sum_m\int_{-\Lambda_x}^{\Lambda_x}dx~\int_{-\Lambda_y}^{\Lambda_y}dy~\frac{1}{(i\omega_m-x-y)}\frac{1}{(i\omega_m-x+y)}\frac{1}{(i\omega_m+x-y)}.
\eeq
It evaluates to,
\beq
Y_{261}\bigg|_1&=&-\frac{T}{8\pi v_xv_y}\sum_m\int_{-\Lambda_x}^{\Lambda_x}dx\frac{\tn{sgn}(\omega_m)}{\omega_m(x-i\omega_m)}=-\frac{T}{8\pi v_xv_y}\sum_{m>0}\int_{-\Lambda_x}^{\Lambda_x}dx\frac{2x}{\omega_m(x^2+\omega_m^2)}=0,\\
Y_{261}\bigg|_2&=&-\frac{T\Lambda_x}{2\pi^2v_xv_y}\sum_{m}\int_{|y|>\Lambda_y}dy\frac{1}{(y-i\omega_m)(y^2+\omega_m^2)}=-\frac{T\Lambda_x}{2\pi^2v_xv_y}\sum_{m}\int_{\Lambda_y}^\infty dy \frac{2i\omega_m}{(y^2+\omega_m^2)^2}=0.\nonumber\\
\eeq
Therefore, we see that both pieces evaluate to zero, when working with the linearized dispersions.

\section{Feynman diagrams for hot-spot theory with a finite curvature}
\label{intcurv}
In this appendix, we provide details for the computation of the same diagrams that were evaluated earlier, but now in the presence of a finite fermi-surface curvature, $\kappa$. We already summarized the results in section \ref{rescu}.

We start with the diagrams contributing to $u_a$. These were already well-behaved in the linearized theory in the $T\rightarrow0$ limit, and hence should continue to be so in the presence of a finite $\kappa$. Let us evaluate them nevertheless. The distinct diagrams contributing to $u_a$ after performing the Matsubara summation are given by,
\beq
\tilde{I}_{1212}=\frac{1}{8\pi^2v_xv_y}\int_{-\Lambda_x}^{\Lambda_x}dx\int_{-\Lambda_y}^{\Lambda_y}dy \bigg[ 2\frac{f(\e_1(x,y))-f(\e_2(x,y))}{(\e_1(x,y)-\e_2(x,y))^3} - \frac{f'(\e_1(x,y))+f'(\e_2(x,y))}{(\e_1(x,y)-\e_2(x,y))^2}\bigg].
\eeq
In the above, $f[...]$ is the fermi-dirac distribution function and we have changed variables to $x=v_x k_x, y=v_y k_y$.  $\tilde{I}_{2525}$ is identical in form to the above with $\e_1\rightarrow\e_5$. We evaluate the above integrals numerically as a function of temperature for different values of $\kappa$ at fixed $\alpha$ and vice versa. We find that $\tilde{I}_{2525}$ identically evaluates to $0$ even in the presence of a finite (but small) curvature (recall that $I_{2525}=0$). The results for $\tilde{I}_{1212}$ alongwith a comparison to the analytical predictions for $I_{1212}$ are shown in fig.\ref{curvI}(a).  

Let us now move onto the diagrams contributing to $w_a$, which were singular ($\sim 1/T$) in our computation with the linearized dispersion. After carrying out the Matsubara summation, the distinct diagrams evaluate to,
\beq
{\tilde{I}}_{2565}=-\frac{1}{4\pi^2v_xv_y}\int_{-\Lambda_x}^{\Lambda_x}dx~\int_{-\Lambda_y}^{\Lambda_y}dy~\bigg[\frac{1}{(\e_2(x,y)-\e_5(x,y))^2}\bigg(\frac{f(\e_2(x,y))}{\e_2(x,y)-\e_6(x,y)} - \frac{f(\e_5(x,y))}{\e_5(x,y)-\e_6(x,y)} \bigg) \nonumber\\
+\frac{1}{(\e_6(x,y)-\e_5(x,y))^2}\bigg(\frac{f(\e_6(x,y))}{\e_6(x,y)-\e_2(x,y)} - \frac{f(\e_5(x,y))}{\e_5(x,y)-\e_2(x,y)} \bigg) \nonumber\\
+ \frac{f'(\e_5(x,y))}{(\e_5(x,y)-\e_2(x,y))(\e_5(x,y)-\e_6(x,y))} \bigg],\nonumber\\ \\
\tilde{I}_{1256}=-\frac{1}{32\pi^2v_xv_y}\int_{-\Lambda_x}^{\Lambda_x}dx~\int_{-\Lambda_y}^{\Lambda_y}dy~\bigg[\frac{f(\e_1(x,y))-f(\e_5(x,y))}{xy(x+y)} - \frac{f(\e_2(x,y))-f(\e_6(x,y))}{xy(x-y)} \bigg],\nonumber\\
\eeq

where we have used the explicit forms of the dispersions to simplify the expression for $\tilde{I}_{1256}$. We evaluate these diagrams numerically and find that both of them have a very similar behavior except at low temperatures, where $\tilde{I}_{2565}$ is always significantly smaller than $\tilde{I}_{1256}$ (this was the case even in our previous computation where the former went as $\sim\log(T)$ while the latter was $\sim 1/T$). We plot $\tilde{I}_{1256}$ in fig.\ref{curvI} (b).

\begin{figure*}
\begin{center}
\includegraphics[width=0.95\columnwidth]{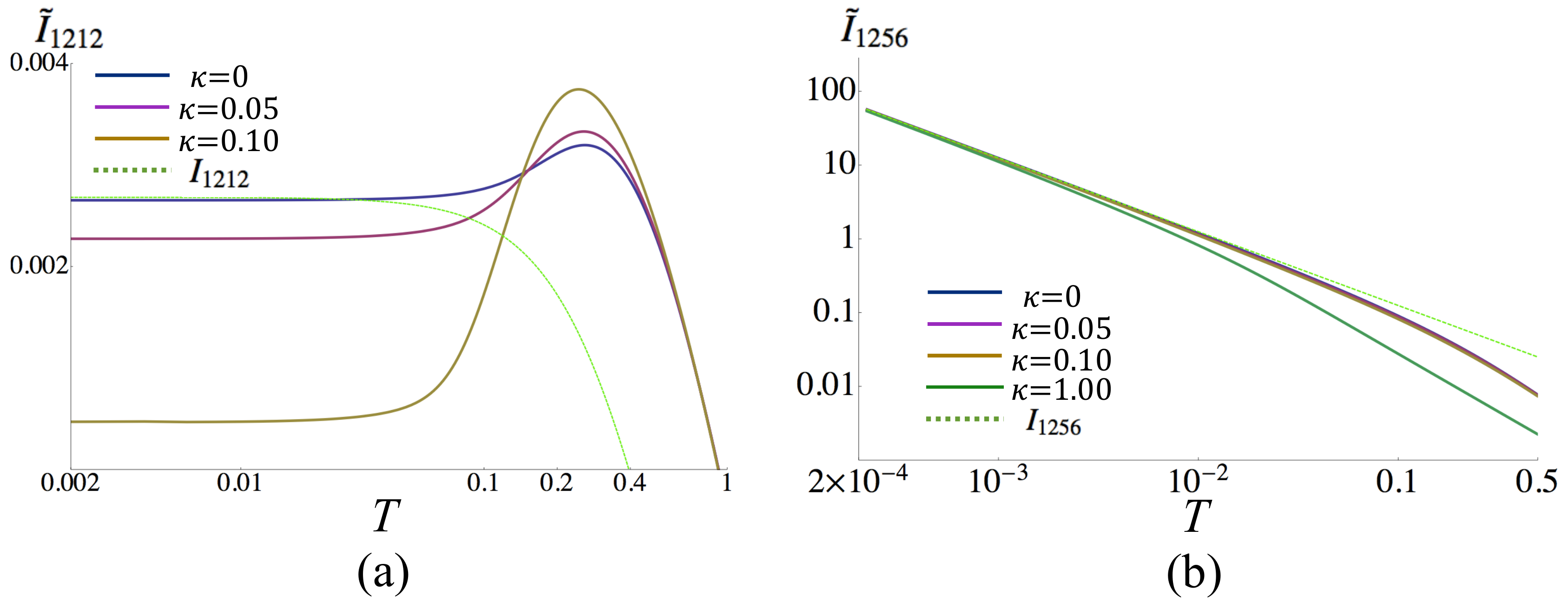}
\end{center}
\caption{Absolute values of diagrams contributing to $u_a$ and $w_a$ as a function of temperature, $T$ for (a), (b), different values of $\kappa$ but fixed $\alpha=10.0$. Other parameters are $\Lambda=2.0$ and $v_x=0.5$. Note the almost perfect agreement of the analytical result for $\kappa=0$ (dashed green line) with the numerical results for small curvature at low temperatures.}
\label{curvI}
\end{figure*}

\begin{figure*}
\begin{center}
\includegraphics[width=1.0\columnwidth]{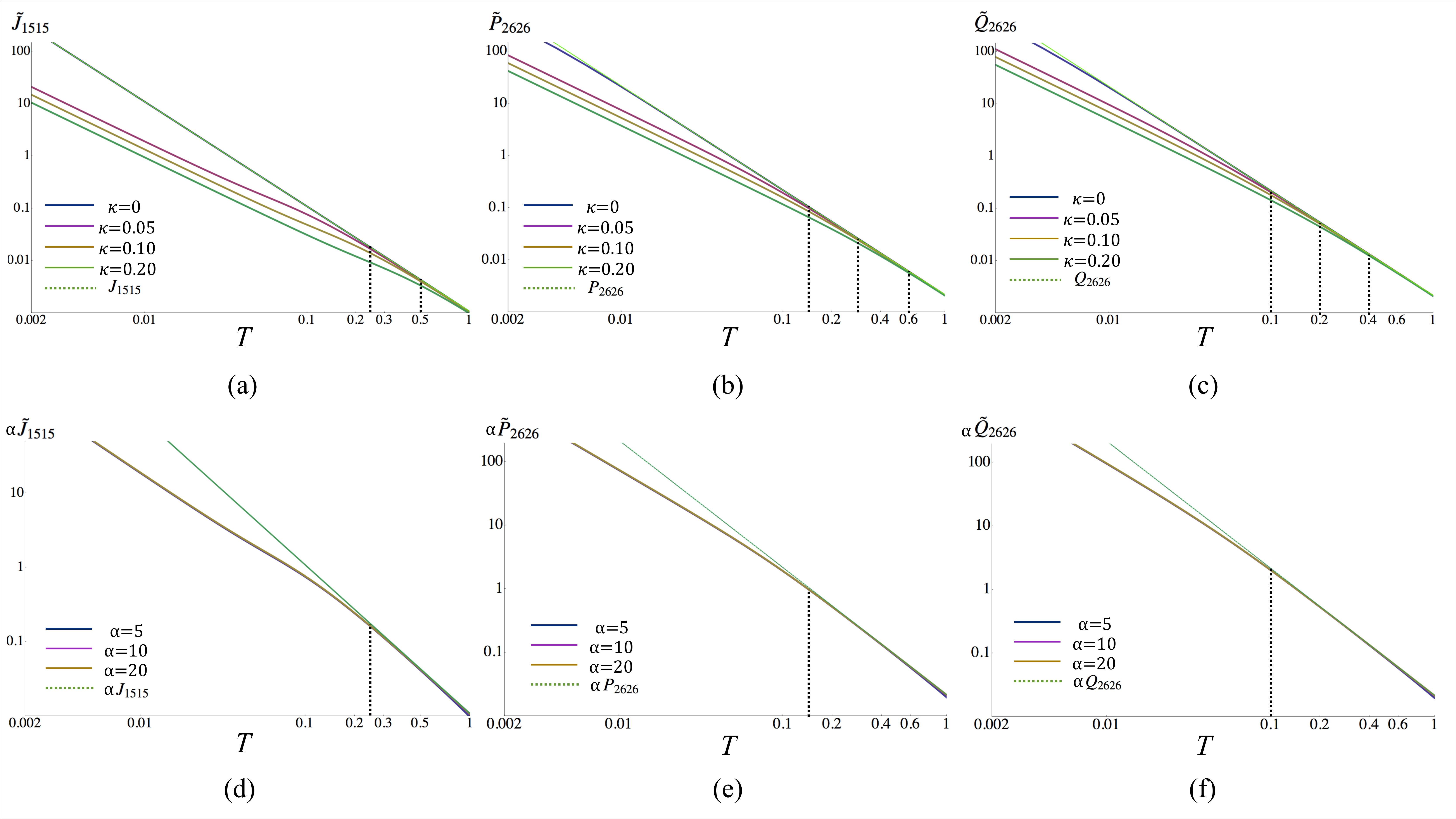}
\end{center}
\caption{Absolute values of diagrams contributing to $u_b$ and $s_b$ as a function of temperature, $T$ for (a), (b), (c) different values of $\kappa$ but fixed $\alpha=10.0$ and (d), (e), (f) different values of $\alpha$ but fixed $\kappa=0.05$.  (a) $\tilde{J}_{1515}$,  (b) $\tilde{P}_{2626}$, (c) $\tilde{Q}_{2626}$. (d)-(f) plot the same diagrams scaled with $\alpha$.   Other parameters are $\Lambda=2.0$ and $v_x=0.5$. The vertical black dotted lines represent the approximate temperature where the computation in the presence of a non-zero $\kappa$ starts deviating from the one with $\kappa=0$. Note that in figs. (a)-(c), the green dashed curves (representing the analytical results) overlap almost perfectly with the blue solid curves for $\kappa=0$.}
\label{curv1t2}
\end{figure*}

Let us now compute all the terms that turned out to be $\sim 1/T^2$ in our earlier computation, which included both $u_b$ ($|\Phi^b|^4$) and $s_b$ ($|\Psi|^2|\Phi^b|^2$),  and study the effect of a finite $\kappa$. The diagrams that contribute to $u_b$ are modified to,
\beq
\tilde{J}_{1515}=\frac{1}{8\pi^2v_xv_y}\int_{-\Lambda_x}^{\Lambda_x}dx~\int_{-\Lambda_y}^{\Lambda_y}dy~\bigg[2\frac{f(\e_1(x,y))-f(\e_5(x,y))}{(\e_1(x,y)-\e_5(x,y))^3} - \frac{f'(\e_1(x,y))+f'(\e_5(x,y))}{(\e_1(x,y)-\e_5(x,y))^2} \bigg],\nonumber\\
\eeq
and $\tilde{J}_{2626}=\tilde{J}_{1515}$ even for $\kappa\neq0$. 

On the other hand, the self-energy type diagrams contributing to $s_b$ are modified to,
\beq
\tilde{P}_{2626}=
 -\frac{1}{4\pi^2v_xv_y}\int_{-\Lambda_x}^{\Lambda_x}dx\int_{-\Lambda_y}^{\Lambda_y}dy\bigg[\frac{f(\e_6(x,y))-f(\e_2(x,y))}{(\e_6(x,y)-\e_2(x,y))(\e_6^2(x,y)-\e_2^2(x,y))}  \nonumber\\
+ \frac{1-2f(\e_6(x,y))}{4\e_6^2(x,y) (\e_2(x,y)+\e_6(x,y))} - \frac{f'(\e_6(x,y))}{2\e_6(x,y)(\e_6(x,y)-\e_2(x,y))}\bigg],
\eeq

and $\tilde{P}_{1515}=\tilde{P}_{2626}$, even when $\kappa\neq0$. Similarly, the vertex-correction diagrams are modified to,
\beq
\tilde{Q}_{2626}&=& -\frac{1}{8\pi^2v_xv_y}\int_{-\Lambda_x}^{\Lambda_x}dx~\int_{-\Lambda_y}^{\Lambda_y}dy~\bigg[\frac{1-2f(\e_6(x,y))}{\e_6(x,y)}  - \frac{1-2f(\e_2(x,y))}{\e_2(x,y)} \bigg]~\frac{1}{\e_2^2(x,y)-\e_6^2(x,y)},\nonumber\\
\eeq
and $\tilde{Q}_{1515}=\tilde{Q}_{2626}$. 
 The results for $\tilde{J}_{1515}$, $\tilde{P}_{2626}$ and $\tilde{Q}_{2626}$ are plotted in fig.\ref{curv1t2}, alongwith a comparison to the respective diagrams evaluated with $\kappa=0$. It is not surprising that the singular power-law agrees, but even the prefactor matches perfectly.

\begin{figure*}
\begin{center}
\includegraphics[width=1.0\columnwidth]{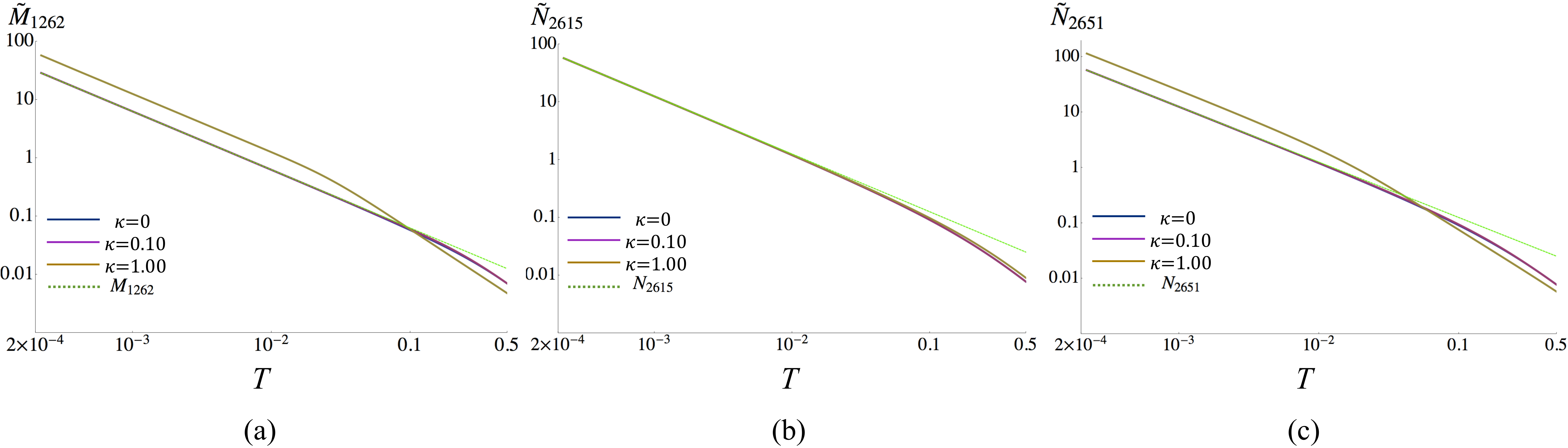}
\end{center}
\caption{Absolute values of diagrams contributing to $s_a$ as a function of temperature, $T$ for (a) $\tilde{M}_{1262}$, (b), $\tilde{N}_{2615}$, (c) $\tilde{N}_{2651}$ for different values of $\kappa$ but fixed $\alpha=10.0$. Other parameters are $\Lambda=2.0$ and $v_x=0.5$. Note the almost perfect agreement of the analytical result for $\kappa=0$ (dashed green line) with the numerical results for small curvature at low temperatures.}
\label{curvMN}
\end{figure*}

Next, we compute the diagrams contributing to $s_a$ ($|\Psi|^2|\Phi^a|^2$). The distinct self-energy type diagrams evaluate to,
\beq
\tilde{M}_{1262}=\frac{1}{4\pi^2v_xv_y}\int_{-\Lambda_x}^{\Lambda_x}dx~\int_{-\Lambda_y}^{\Lambda_y}dy~\bigg[\frac{f(\e_1(x,y))-f(\e_2(x,y))}{(\e_1(x,y)-\e_2(x,y))(\e_1^2(x,y)-\e_2^2(x,y))}  \nonumber\\
- \frac{1-2f(\e_2(x,y))}{4\e_2^2(x,y)(\e_2(x,y)+\e_1(x,y))} + \frac{f'(\e_2(x,y))}{2\e_2(x,y)(\e_2(x,y)-\e_1(x,y))}\bigg]
\eeq

and $\tilde{M}_{2515}$ is identical in form to the above with the replacement, $\e_1\rightarrow\e_2$ and $\e_2\rightarrow\e_5$.

Similarly, the distinct vertex-correction type diagrams evaluate to,
\beq
\tilde{N}_{2615}= -\frac{1}{8\pi^2v_xv_y}\int_{-\Lambda_x}^{\Lambda_x}dx~\int_{-\Lambda_y}^{\Lambda_y}dy~\bigg[\frac{1-2f(\e_2(x,y))}{\e_2(x,y)} 
- \frac{1-2f(\e_5(x,y))}{\e_5(x,y)} \bigg]~\frac{1}{\e_5^2(x,y)-\e_2^2(x,y)},\nonumber\\
\label{N}
\eeq 
and where $\tilde{N}_{2651}$ can be obtained from the above by replacing $\e_5\rightarrow\e_1$. The results are plotted in fig.\ref{curvMN}.

Finally, let us evaluate the three-point functions, $t_{ab}$. Recall that in the linearized theory, this was identically $0$. In the presence of a curvature, it is modified to,
\beq
\tilde{Y}_{261}=-\frac{1}{4\pi^2v_xv_y}\int_{-\Lambda_x}^{\Lambda_x}dx \int_{-\Lambda_y}^{\Lambda_y}dy \bigg[\frac{f(\e_1(x,y))}{(\e_1(x,y)-\e_2(x,y))(\e_1(x,y)-\e_6(x,y))} \nonumber\\
+ \frac{1}{\e_2(x,y)-\e_6(x,y)}\bigg(\frac{f(\e_2(x,y))}{\e_2(x,y)-\e_1(x,y)}-\frac{f(\e_6(x,y))}{\e_6(x,y)-\e_1(x,y)} \bigg) \bigg],
\eeq
and where $\tilde{Y}_{261}$ is still equal to the other symmetry related diagrams. The results for $\tilde{Y}_{261}/\kappa$ and $\alpha\tilde{Y}_{261}$ are shown in figs.\ref{y261} (a) and (b) respectively, alongwith a comparison to the particular form, ${\mathcal{Y}}$, that we guessed.  
\begin{figure}
\begin{center}
\includegraphics[width=0.9\columnwidth]{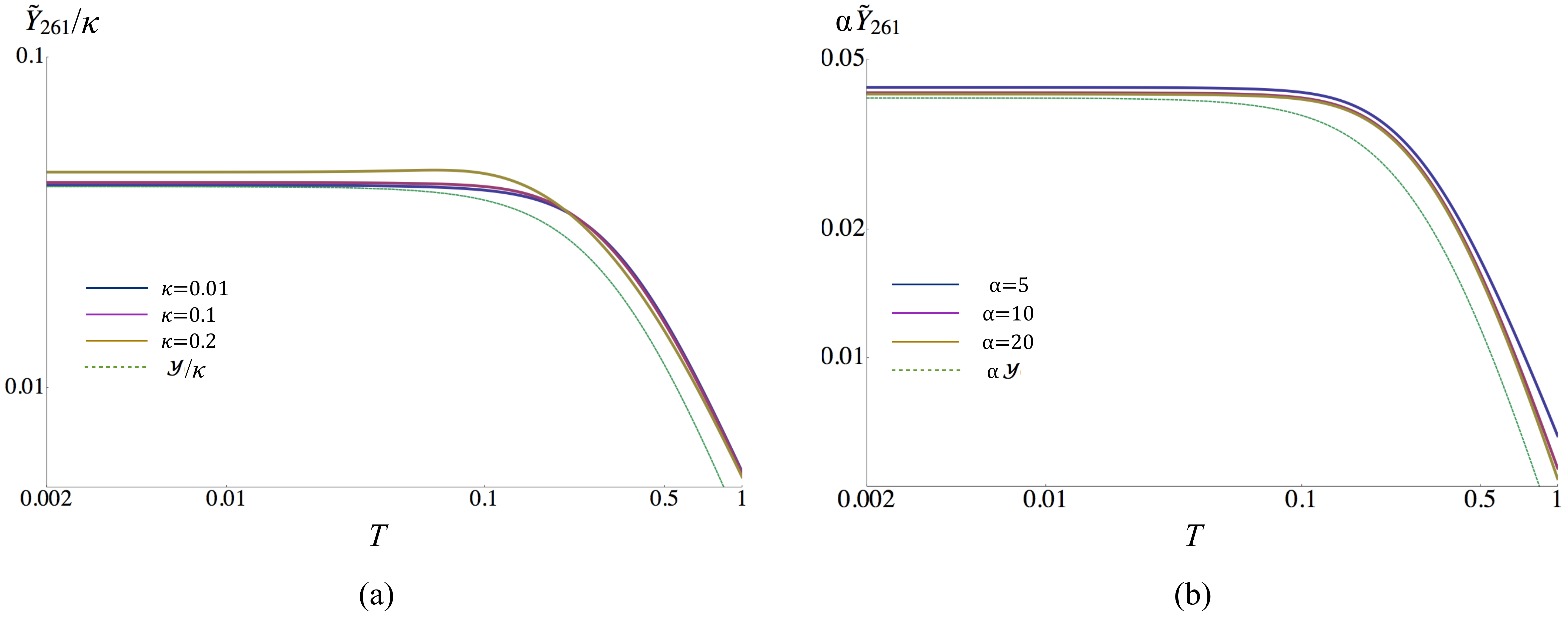}
\end{center}
\caption{Absolute value of $\tilde{Y}_{261}$ as a function of temperature and comparison with ${\cal{Y}}$ (a) at fixed $\alpha=10$, (b) at fixed $\kappa=0.1$}
\label{y261}
\end{figure}

\end{widetext}


\begin{thebibliography}{99}
\bibitem{commenttemp} However, we do not address here the issue of how the onset temperature for CDW-a compares to the bare superconducting $T_c$.
\end{thebibliography}
\end{document}